\documentclass[authoryear,preprint,3p,times,11pt]{elsarticle}

\usepackage{tikz}
\usepackage{txfonts}
\usepackage{natbib}
\usepackage{epsfig}
\usepackage{float}
\usepackage{graphicx}
\usepackage{epstopdf}
\usepackage{amssymb}
\usepackage[colorlinks,linkcolor=blue,hyperindex,CJKbookmarks]{hyperref}
\usepackage{bm}
\usepackage{amsmath}
\usepackage{pdflscape}
\usepackage{array}
\usepackage[english]{babel}
\usepackage[font=small,labelfont=bf,tableposition=top]{caption}
\usepackage{booktabs}
\usepackage{threeparttable}
\usepackage{lscape}
\usepackage{subfigure}
\usepackage{multirow}
\usepackage{upgreek}
\usepackage{comment}
\usepackage{setspace}
\usepackage{appendix}
\usepackage{mathtools}
\usepackage{cases}
\usepackage{xcolor}
\usepackage{hhline}
\usepackage{makecell}
\usepackage{soul}
\usepackage{ntheorem}
\usepackage{enumerate}
\usepackage{ulem}
\usepackage[displaymath,mathlines,edtable]{lineno}
\usepackage{lipsum}
\usepackage{cuted}
\usepackage[margins]{misc/trackchanges}
\usepackage{booktabs}
\usepackage{longtable}
\usepackage{array}
\usepackage{cleveref}

\usepackage{etoolbox} 

\usepackage[section]{placeins}

\usepackage[ruled]{algorithm2e} 

\newcommand{\EQ}{\begin{eqnarray}}
\newcommand{\EN}{\end{eqnarray}}
\newcommand{\EQQ}{\begin{eqnarray*}}
\newcommand{\ENN}{\end{eqnarray*}}

\newcommand{\bremark}{\begin{remark} \begin{rm}}
\newcommand{\eremark}{ \end{rm} \rule{1mm}{2mm}
\end{remark}}
\newcommand{\basm}{\begin{assumption} \begin{rm}}
\newcommand{\easm}{\end{rm}
\end{assumption}}
\newcommand{\bsup}{\begin{supposition} \begin{rm}}
\newcommand{\esup}{\end{rm}
\end{supposition}}
\newcommand{\btheorem}{\begin{theorem} \begin{rm}}
\newcommand{\etheorem}{\end{rm} \rule{1mm}{2mm}
\end{theorem}}
\newcommand{\blemma}{\begin{lemma}\begin{rm}}
\newcommand{\elemma}{\end{rm} \rule{1mm}{2mm}
\end{lemma}}
\newcommand{\bcorollary}{\medskip\begin{corollary}\begin{rm}}
\newcommand{\ecorollary}{\end{rm}\rule{1mm}{2mm}
\end{corollary}}
\newcommand{\bdefinition}{\medskip\begin{definition}\begin{rm}}
\newcommand{\edefinition}{\end{rm}\rule{1mm}{2mm}
\end{definition}}
\newcommand{\bproposition}{\medskip\begin{proposition}\begin{rm}}
\newcommand{\eproposition}{\end{rm}\rule{1mm}{2mm}
\end{proposition}}
\newcommand{\bexample}{\begin{example}\begin{rm}}
\newcommand{\eexample}{\end{rm}\rule{1mm}{2mm}\end{example}}

\newcommand{\bcase}{\begin{case}\begin{rm}}
\newcommand{\ecase}{\end{rm}\rule{1mm}{2mm}\end{case}}

\newtheorem{theorem}{\bf Theorem}[section]
\newtheorem{lemma}{\bf Lemma}[section]
\newtheorem{definition}{\bf Definition}[section]
\newtheorem{remark}{\bf Remark}[section]
\newtheorem{corollary}{\bf Corollary}[section]
\newtheorem{proposition}{\bf Proposition}[section]
\newtheorem{example}{\bf Example}[section]
\newtheorem{assumption}{\bf Assumption}[section]

\newtheorem{case}{\bf Case}[section]
\newtheorem{supposition}{\bf Hypothesis}[section]

\def\equationautorefname~#1\null{(#1)\null}

\graphicspath{{Figure/}}

\journal{}

\begin{document}
\begin{sloppypar}  

\begin{frontmatter}
\title{Calibration and uncertainty quantification of macroscopic fundamental diagrams}

\author[sysu]{Wenfei Ma}
\ead{mawf5@mail2.sysu.edu.cn}
\author[calgary]{Yunping Huang}
\ead{yunping.huang@connect.polyu.hk}
\author[Monash]{Nan Zheng}
\ead{nan.zheng@monash.edu}
\author[pc]{Tianlu Pan\corref{cor1}}
\ead{pantl@pcl.ac.cn}
\author[sysu]{Renxin Zhong\corref{cor1}}
\ead{zhrenxin@mail.sysu.edu.cn}

\cortext[cor1]{Corresponding author.}
\address[sysu]{
School of Intelligent Systems Engineering,  Sun Yat-Sen University (Shenzhen Campus), Guangdong, China.}
\address[calgary]{Department of Civil and Environmental Engineering, Schulich School of Engineering, University of Calgary, Calgary, Alberta, Canada.}
\address[Monash]{Institute of Transport Studies, Department of Civil Engineering, Monash University, Melbourne, Australia.}
\address[pc]{Department of Network Intelligence, Peng Cheng Laboratory, Shenzhen, China.}

\begin{abstract}
Traffic congestion occurs as travel demand exceeds network capacity, necessitating a thorough understanding of network capacity for effective traffic control and management. The macroscopic fundamental diagram (MFD) provides an efficient framework for quantifying network capacity. However, empirical MFDs exhibit considerable data scatter and uncertainty. In this paper, we propose a mathematical program that simultaneously calibrates the MFD and quantifies the uncertainty associated with data scatter. We further investigate contributing factors of uncertainties regarding network capacity and traffic resilience. To be specific, we first include two conventional approaches for MFD calibration and uncertainty quantification as special cases. The proposed program is validated using empirical data from two cities in China. Subsequently, we identify how congestion loading and recovery contribute to data scatter. We develop a novel uncertainty quantification approach capable of capturing distinct congestion phases simultaneously. The gap between the upper and lower bounds of the calibrated MFDs, represented by a coefficient parameter, is regarded as the capacity drop induced by traffic congestion. Furthermore, the proposed approach enables further exploration of how macroscopic factors, such as travel demand and traffic control, and microscopic factors, such as driving behavior, influence MFD hysteresis, uncertainty, and traffic resilience. These findings shed light on developing efficient traffic control and demand management strategies to increase network capacity and resilience, while we count on future connected automated vehicular technologies to conquer the influence of various driving behaviors.

\end{abstract}


\end{frontmatter}


\section{Introduction} \label{sec:MFD:cali:intro}
Traffic congestion is a long-standing challenge that arises from the imbalance between travel demand and network capacity. To effectively design traffic control and demand management strategies, it is crucial to have a thorough understanding of network capacity. 
The macroscopic fundamental diagram (MFD) provides a framework for quantifying network capacity, offering insights into the network supply function. MFDs describe the relationships between average network flow and density, as well as the trip completion rate and accumulation under stationary conditions. These network-wide relationships have been studied since the 1960s \citep{Godfry1969the, herman1979two, mahmassani1984investigation}, and received renewed interest recently since the well-defined relationships were validated empirically \citep{geroliminis2008existence}. 
MFDs provide a straightforward analytical framework with computational efficiency to model large-scale urban network dynamics.
This efficient framework has been validated in various traffic applications, including urban network traffic control \citep{haddad2012stability, aboudolas2013perimeter, Keyvan2015a, zhong2018robust, zhong2018boundary, su2020neuro, chen2022data, su2023hierarchical, chen2024iterative}, traffic state estimation \citep{Ambuhl2016data, Kouvelas2017real, Mohammadreza2021an}, dynamic traffic assignment \citep{yildirimoglu2013dynamic, yildirimoglu2014approximating, huang2020dynamic,zhong2020dso, zhong2021dynamic}, and vehicle dispatching \citep{ramezani2018dynamic, alisoltani2020sequential, beojone2021on, Valadkhani2023dynamic, huang2024bi}.

Numerous studies based on empirical or simulated data have been conducted to calibrate MFDs of urban networks worldwide.
\cite{geroliminis2008existence} initially validated the existence of MFD linking space-mean flow, density, and speed based on loop detector data and floating car data from Yokohama, Japan. Subsequently, \cite{buisson2009Exploring} investigated the flow-occupancy MFD using loop detector data of Toulouse, France. \cite{cassidy2011macroscopic} explored the MFD that related flow to density, or equivalently, vehicle kilometers traveled (VKT) to vehicle hours traveled (VHT) of the freeway network located in California, USA. \cite{Ambuhl2016data} described the MFD relating average flow and density with loop detector data and floating car data of Zurich, Switzerland.
These MFD estimation studies generally assume a well-defined deterministic relationship. 
However, MFDs derived from empirical or simulated data exhibit considerable scatter and uncertainty, indicating that a range of productivities may be observed for a given network accumulation.
For instance, in the MFD originally presented for Yokohama, Japan, the observed average network flows exhibit fluctuations of up to 10\% around the mean value at certain densities \citep{geroliminis2008existence}. 
While most MFD calibration approaches rely on a deterministic mapping between network accumulation and productivity, incorporating uncertainties in MFDs is crucial for enhancing the modeling of traffic dynamics \citep{haddad2016adaptive}. 
Furthermore, these uncertainties cannot be eliminated \citep{gao2018analytical}.

The capability of deterministic MFDs to accurately represent the relationship between network accumulation and productivity may be limited without considering the scatter and uncertainty. Consequently, traffic control and demand management strategies derived from MFD-based dynamics might be inaccurate or misleading. On the other hand, the implementation of these strategies can influence MFD shape and uncertainty in turn. 
Therefore, the investigation and quantification of uncertainty within MFDs emerge as a prominent research topic.
Recent studies confirm that inherent MFD uncertainties can lead to fundamentally different network dynamics, such as extreme congestion or gridlock \citep{gao2018analytical}. 
Stochastic MFD models have been proposed to capture these uncertainties, treating the critical parameters as random variables \citep{ he2024learning}. In the broader realm of fundamental diagrams (FDs), studies have also endeavored to model and analyze the uncertainties embedded in the real-world data \citep{shi2021physics, bai2021calibration, kriuchkov2023stochastic,cheng2023analytical}.
However, most studies primarily focus on incorporating MFD uncertainty into traffic control and management strategies (e.g., perimeter control and tolling), with limited consideration of the interactive effects between MFD uncertainty and these strategies. 
It is essential to develop a systematic framework for MFD uncertainty quantification and further investigate its implications for dynamic traffic control and management.

The scatter and uncertainty of MFD is found to be influenced by travel demand \citep{paipuri2019validation, niu2022impact, xu2023opposing}, traffic control \citep{keyvan2019traffic, alonso2019effects, su2023hierarchical, de2024evaluation}, driving behavior \citep{mahmassani2013urban, lu2020impact}, and network topology \citep{buisson2009Exploring, geroliminis2011Hysteresis, Saberi2012exploring, saberi2013hysteresis, knoop2015network}.
Studies have revealed that MFD with peak demand exhibits significant scatter, known as the hysteresis loop, and increased demand variability leads to more pronounced hysteresis patterns \citep{paipuri2019validation, xu2023opposing}. Analytical frameworks have been developed to model the traffic dynamics considering travel demand uncertainty \citep{gao2018analytical}.
Traffic control is another crucial factor contributing to MFD uncertainty. Research indicates that effective signal control increases network capacity \citep{zhang2013comparative}, 
and the coordination of adaptive signal control and perimeter control further enhances the capacity \citep{keyvan2019traffic, su2023hierarchical, de2024evaluation}. Robust control strategies incorporating uncertainties within bounded parameters have been proposed for MFD networks \citep{haddad2015robust, zhong2018robust}. 
Furthermore, recent efforts have increasingly focused on the impact of driving behaviors on MFD uncertainty. Studies suggest that a higher proportion of adaptive drivers, who adjust their routes based on real-time traffic conditions, reduces the probability of the occurrence of MFD hysteresis \citep{daganzo2011macroscopic, gayah2011clockwise, mahmassani2013urban}. 
Additionally, studies on automated vehicles (AVs) reveal that network capacity improves with a higher penetration rate of AVs and shorter AV headway settings \citep{lu2020impact, shi2021constructing}. 
Network topology has also been shown to significantly affect the shape of MFDs \citep{buisson2009Exploring, geroliminis2011Hysteresis, Saberi2012exploring, saberi2013hysteresis}. 
Research indicates that networks with different topologies exhibit different levels of data scatter and network capacity \citep{ji2010Investigating, knoop2015network}.
Despite the recognized impact of these factors on MFD uncertainty, the extent and mechanisms of these effects, as well as their implications for traffic control and management, remain insufficiently explored. This study therefore investigates how demand-side and supply-side factors influence MFD uncertainty, with particular attention to their implications for traffic control and management strategies.

\begin{figure}[!htbp]
\centering
\includegraphics[width=1\linewidth]{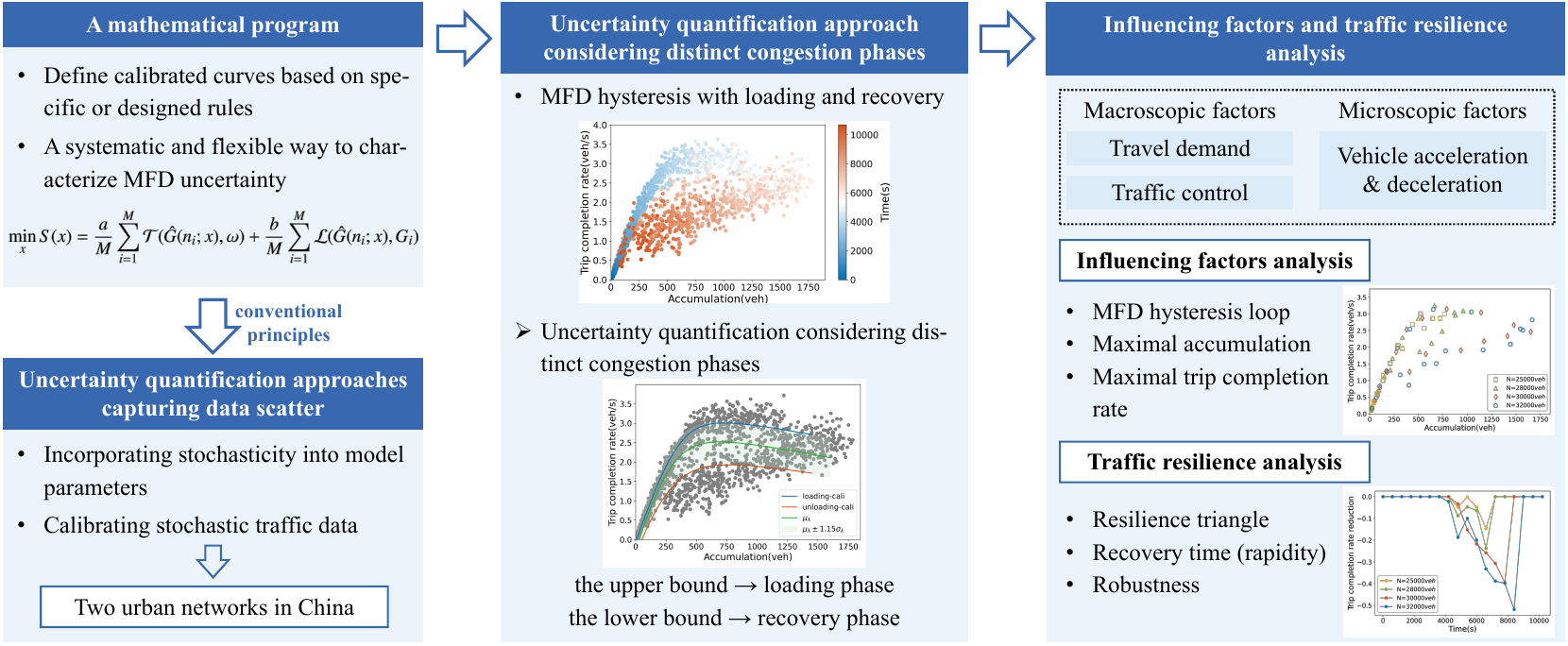}
\caption{Framework for MFD calibration and uncertainty quantification.}
\label{fig:frame}
\end{figure}

Additionally, MFD provides a straightforward way to evaluate network performance without detailed traffic physics. By integrating MFD dynamics and network functionality, reliable traffic resilience indicators have been developed \citep{kim2017evaluating, amini2018evaluating, lu2024traffic}. 
Resilience loss under congestion occurs when network accumulation exceeds its critical value, preventing optimal traffic operations \citep{lu2024traffic}, and further reducing network capacity \citep{nogal2016resilience}.
Given that the trip completion rate indicates the service level of traffic networks, resilience loss under congestion can be assessed by measuring the decrease in trip completion rate \citep{lu2024traffic}.
Moreover, MFDs with uncertainty usually exhibit significant capacity reduction, particularly when the hysteresis phenomenon is present. 
It is essential to explore the relationship between MFD uncertainty and traffic network performance, with MFD-based resilience indicators providing an intuitive framework for this analysis.
To fill the above research gaps, we propose an MFD calibration and uncertainty quantification framework, as shown in \autoref{fig:frame}.
The primary contributions of this study are as follows:
\begin{itemize}
    \item We propose a unified mathematical program simultaneously for MFD calibration and uncertainty quantification, and further include conventional quantification principles within the program as special cases, presenting two approaches for capturing MFD data scatter. The program is validated using empirical data from two urban networks in China.  
    \item Motivated by the observed MFD hysteresis, we infer that different congestion loading and recovery physics contribute to data scatter and develop a novel uncertainty quantification approach based on the proposed mathematical program for MFD calibration. Validated through microscopic simulation, the approach effectively captures distinct congestion phases, with a coefficient parameter indicating capacity drop under traffic congestion.    
    \item We investigate the effects of macroscopic factors, including travel demand and traffic control, and microscopic factors, including vehicle acceleration and deceleration, on network capacity and MFD uncertainty. An MFD-based resilience indicator is employed to evaluate the impact of hysteresis, demonstrating that networks exhibiting lower uncertainty are more resilient to congestion disruptions.
\end{itemize}

The rest of this paper is organized as follows: 
\autoref{sec:MFD:cali:preliminaries} presents a mathematical program for MFD calibration and uncertainty quantification, using the $\lambda$-trapezoidal MFD function as an illustrative example.
\autoref{sec:MFD:cali:uncertainty} includes two conventional quantification principles as special cases, and validates the program using data from two urban networks.
\autoref{sec:MFD:cali:quantification} proposes a novel uncertainty quantification approach that accounts for distinct congestion loading and recovery phases, and validates it through a microscopic simulation.
\autoref{sec:MFD:cali:influence_resilience} analyzes the mechanisms by which factors influence network capacity and MFD uncertainty, and explores their relationship with traffic resilience.
Finally, \autoref{sec:MFD:cali:conclusion} concludes the paper with future directions.
\section{A mathematical program for MFD calibration and uncertainty quantification}
\label{sec:MFD:cali:preliminaries}

MFD calibration aims to derive representative curves that best fit the observed data
to capture the aggregated traffic dynamics.
However, the scatter of MFD data suggests that a single calibrated curve may inadequately capture the variability of network dynamics.
Quantifying MFD uncertainty is thus essential to account for the scatter and ensure a robust representation of traffic dynamics \citep{haddad2016adaptive}.
Generally, the core of uncertainty quantification is to define multiple curves that either enclose the data scatter or follow specific theoretical principles. 
For example, \cite{cheng2023analytical} developed stochastic fundamental diagram models based on a probabilistic framework to reproduce the observed scatter and explain traffic state variations.


Building on these perspectives, we propose the following mathematical program for MFD calibration that generates one or more calibrated curves by specifying their location. 
\begin{align}
    \min_x S(x) =\frac{a}{M}\sum_{i=1}^M \mathcal{T}(\hat{G}(n_i;x),\omega)+ \frac{b}{M} \sum_{i=1}^M \mathcal{L}(\hat{G}(n_i;x),G_i)
    \label{eq:general_quantify}
\end{align}
where $(n_i, G_i) \in D^{M}$ is the observed data in dataset $D^M$. 
$\hat{G}(n;x)$ denotes the calibrated MFD curve parameterized by $x$, and $\mathcal{T}(\cdot)$ is a function with a predefined parameter $\omega$ that defines the location of the curve based on specified rules or desired properties. 
The loss function $\mathcal{L}(\cdot)$ quantifies the deviation between the observed $G_i$ and the output of the MFD curve $\hat{G}(n_i;x)$ to ensure it remains close to the data points. 
The coefficients $a$ and $b$ balance the trade-off between curve localization and fitting accuracy.
These curves, derived under different fitting principles or location constraints, characterize distinct aspects of data scatter. 
Consequently, this program provides a unified framework for simultaneously calibrating MFD and quantifying its uncertainty.

In the absence of location constraints (i.e., omitting the first term), the formulation reduces to a standard calibration problem. 
Specifically, when the loss function is defined as the mean absolute percentage error (MAPE), the formulation aligns with Eq.(8) in \cite{ma2024functional}. Furthermore, when $\mathcal{L}(\cdot)$ is designed as the squared differences between the observed and curve values, the formulation turns into the widely applied least squares (LS) method for model calibration \citep{may1990traffic}.



\begin{figure}[!htbp]
  \centering
  \includegraphics[width=0.5\linewidth]{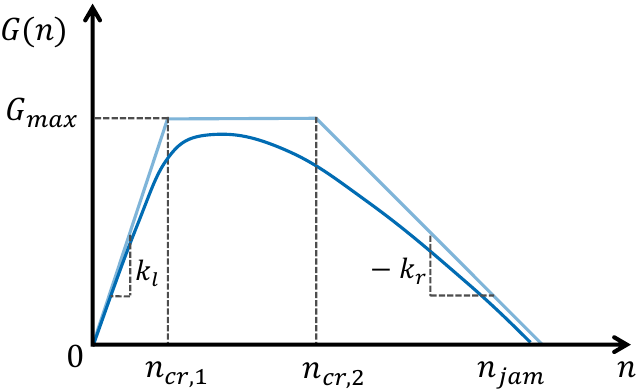}\\
  \caption{The $\lambda$-trapezoidal MFD with its upper bound.}
  \label{fig:linear_MFD}
\end{figure}

Our previous work \citep{ma2024functional} proposed a two-step MFD calibration framework that automatically selects the most appropriate functional form from candidate functions. Similarly, the mathematical program proposed in this paper is applicable to various MFD functional forms. Without loss of generality, we demonstrate the mathematical program and subsequent uncertainty quantification approaches using the $\lambda$-trapezoidal MFD \citep{ambuhl2020afunctional}. The $\lambda$-trapezoidal MFD offers physical interpretations for each parameter and has been empirically validated using data from 41 cities worldwide, encompassing 107 regional networks \citep{loder2019understanding}.
The $\lambda$-trapezoidal MFD is formulated as:
\begin{subequations}
\begin{align}
     G(n) & =- \lambda \ln 
    \left(
    e^{-\frac{k_l n}{\lambda}}+e^{-\frac{G_{max}}{\lambda}}+e^{-\frac{k_r(n_{jam}-n)}{\lambda}}
    \right)
    \label{eq:lambda}
    \\
    G_{up}(n) &=\left \{
    \begin{aligned}
        & k_ln, & n\leq n_{cr,1};\\
        & G_{max}, & n_{cr,1} < n \leq n_{cr,2};\\
        & k_r(n_{jam}-n), & n>n_{cr,2}. 
    \end{aligned}
    \right.
    \label{eq:lambda_bound}
\end{align}
\end{subequations}
where $G_{max}$ denotes the critical trip completion rate and $n_{jam}$ denotes the jam accumulation. 
The upper bound of the $\lambda$-trapezoidal MFD, denoted by $G_{up}(\cdot)$, is defined in \eqref{eq:lambda_bound}.
The slopes of $G_{up}(\cdot)$ are defined as $k_l=G_{max}/n_{cr,1}$, $k_r=G_{max}/(n_{jam}-n_{cr,2})$, as depicted in \autoref{fig:linear_MFD}. 
The parameter vector of upper bound $G_{up}(\cdot)$ denotes as $x_{up}=[k_l,k_r, G_{max},n_{jam}]$. 
The parameter $\lambda$ is a smoothing factor that approximates the curve from its upper bound. As defined by \eqref{eq:lambda}, an increase in $\lambda$ leads to a decrease in trip completion rates. Therefore, $\lambda$ can be interpreted as the capacity reduction caused by infrastructure effects and vehicle interactions \citep{ambuhl2020afunctional}.
The $\lambda$-trapezoidal MFD is characterized by the parameter vector $x=[k_l,k_r, G_{max},n_{jam},\lambda]$.

\section{Uncertainty quantification approaches capturing data scatter} 
\label{sec:MFD:cali:uncertainty}

In this section, we present two approaches for capturing the scatter in MFD data, integrating conventional quantification principles as special cases within the proposed mathematical program.
The first approach introduces stochasticity into model parameters to capture MFD uncertainty, while the second approach directly calibrates the stochastic traffic data to characterize scatter.
Without loss of generality, both approaches are demonstrated using the $\lambda$-trapezoidal MFD \citep{ambuhl2020afunctional}.
Empirical data from two urban networks in China are used to validate their performance.

\subsection{Uncertainty quantification by incorporating stochasticity into parameters} \label{sec:MFD:cali:uncertainty_1}

To capture the data scatter in MFD, parameters can be modeled as random variables \citep{cheng2023analytical, he2024learning}. However, it inherently increases model complexity. It is crucial to randomize as few parameters as possible while effectively capturing the data scatter.
%
In the $\lambda$-trapezoidal MFD, parameters $x_{up}=[k_l, k_r, G_{max}, n_{jam}]$ define the upper bound of macroscopic data and are relatively stable. 
Therefore, we propose to incorporate stochasticity into the smoothing parameter $\lambda$, assuming it follows a Gaussian distribution. 
The resulting MFD function is formulated as follows:
\begin{align}
    G(n)=- \lambda \ln (e^{-\frac{k_l n}{\lambda}}+e^{-\frac{G_{max}}{\lambda}}+e^{-\frac{k_r(n_{jam}-n)}{\lambda}}), \ \lambda \sim N(\mu_{\lambda},\sigma_{\lambda})
    \label{eq44}
\end{align}
where $N(\mu_{\lambda},\sigma_{\lambda})$ represents a Gaussian distribution, with $\mu_{\lambda}$ denoting the mean and $\sigma_{\lambda}$ denoting the variance. 

Calibrating the $\lambda$-trapezoidal MFD involves two steps: 1) calibrating parameters $x_{up}$ of its upper bound, and 2) calibrating the smoothing parameter $\lambda$.
Based on the proposed mathematical program shown in \eqref{eq:general_quantify}, the objective function to calibrate its upper bound is defined as follows:
\begin{align}
    \min_{x_{up}} S(x_{up}) = \frac{a}{M}\sum_{i=1}^{M} T(n_i;x_{up}) + \frac{b}{M} \sum_{i=1}^{M} |\frac{G_{up}(n_i;x_{up})-G_i}{G_i}  \bigr |
    \label{eq41}
\end{align} 
subject to 
\begin{align}
    T(n_i;x_{up}) = \left \{
    \begin{aligned}
        0, \ & \text{if} \ G_i \leq G_{up}(n_i;x_{up});\\ 
        1, \ & \text{otherwise}.
    \end{aligned}
    \right.
    \label{eq42}
\end{align}
where $T(n_i;x_{up})$ is an indicator function defined in \eqref{eq42} to indicate whether the data points lie below the calibrated curve, serving as a specific case of $\mathcal{T}(\cdot)$ in program \eqref{eq:general_quantify}.
The first term of the objective function aims to position the resulting curve predominantly above the data points, and the second term minimizes the distance between the curve and MFD data, using MAPE as the loss function $\mathcal{L}(\cdot)$.

Given that the smoothing parameter $\lambda$ is assumed to follow a Gaussian distribution, we directly calibrate its distributional parameters, $\mu_{\lambda}$ and $\sigma_{\lambda}$, in the second step.
Define $\lambda_u=\mu_{\lambda}+\sigma_{\lambda}$ and $\lambda_l=\mu_{\lambda}-\sigma_{\lambda}$ to represent the upper and lower bounds of $\lambda$, respectively. Based on these bounds, two MFDs are derived, as shown in \eqref{eq_ealow} and \eqref{eq_eaup}. Notably, due to the definition of parameter $\lambda$, the MFD corresponding to the lower $\lambda$ lies upper.
\begin{subequations}
\begin{align}
    G_l(n)=- \lambda_u\ln (e^{-\frac{k_l n}{\lambda_u}}+e^{-\frac{G_{max}}{\lambda_u}}+e^{-\frac{k_r(n_{jam}-n)}{\lambda_u}})
    \label{eq_ealow}
    \\
    G_u(n)=- \lambda_l\ln (e^{-\frac{k_l n}{\lambda_l}}+e^{-\frac{G_{max}}{\lambda_l}}+e^{-\frac{k_r(n_{jam}-n)}{\lambda_l}})
    \label{eq_eaup}
\end{align}    
\end{subequations}

Leveraging the empirical properties of the Gaussian distribution, the objective function to calibrate $\mu_\lambda$ and $\sigma_\lambda$ is formulated based on the proposed mathematical program as follows:
\begin{align}
    \min_{\mu_\lambda, \sigma_\lambda} S(\lambda)=\frac{a}{M}\sum_{i=1}^M |T_1(n_i;\lambda_u,x_{up})-s|+\frac{b}{M}\sum_{i=1}^M |T_2(n_i;\lambda_l,x_{up})-s|+\frac{c}{M} \sum_{i=1}^M \bigl |\frac{G(n_i;\mu_{\lambda},x_{up})-G_i}{G_i}  \bigr |
    \label{eq47}
\end{align}
subject to
\begin{subequations}
\begin{align}
    \lambda_u & = \mu_{\lambda}+  \sigma_{\lambda}, \ \lambda_l=\mu_{\lambda}-\sigma_{\lambda} 
    \label{eq_lambda_ul}
    \\
    T_1(n_i;\lambda_u,x_{up}) & = \left \{
    \begin{aligned}
        0, \ & if \ G_i \geq G(n_i;\lambda_u,x_{up}); \\
        1, \ & otherwise.
    \end{aligned}
    \right.
    \label{eq_T1}
    \\
    T_2(n_i;\lambda_l,x_{up}) & = \left \{
    \begin{aligned}
        0,\ & if \ G_i \leq G(n_i;\lambda_l,x_{up}); \\
        1,\ & othewise.
    \end{aligned}
    \right.
    \label{eq_T2}
\end{align}
\end{subequations}
where $T_1(n_i;\lambda_u,x_{up})$ and $T_2(n_i;\lambda_l,x_{up})$ are indicator functions defined by \eqref{eq_T1} and \eqref{eq_T2}, respectively.
The parameter $s= 0.16 $ is determined based on the empirical rule of the standard normal distribution, where approximately 68\% of the data lies within $\pm1$ standard deviation of the mean value.
The coefficients $a$, $b$, and $c$ represent the weights of different objectives.
The first two terms in \eqref{eq47} adjust $\sigma_{\lambda}$ by scaling data points positioned between the upper and lower bounds, enabling the framework to accommodate asymmetric distributions. The third term calibrates the mean value $\mu_{\lambda}$ using MAPE as the loss function.

\subsection{Uncertainty quantification by calibrating stochastic traffic data} \label{sec:MFD:cali:uncertainty_2}

The uncertainty quantification method that incorporates stochasticity into parameters assumes that the selected parameters follow specific distributions (e.g., Gaussian distribution) based on prior analysis or empirical observations. However, these predefined choices may not fully capture the characteristics of data scatter. 
To overcome this limitation, we propose directly calibrating stochastic data with a segmentation parameter $\eta$, which allows for a more flexible representation of the data scatter. 
Based on the proposed mathematical program in \eqref{eq:general_quantify}, the objective function is formulated as follows:
\begin{align}
    \min_x S(x) =\frac{a}{M}\sum_{i=1}^M |T(n_i;x)-\eta|+ \frac{b}{M} \sum_{i=1}^M \bigl |\frac{G(n_i;x)-G_i}{G_i}  \bigr |
    \label{eq50}
\end{align} 
subject to
\begin{align}
    T(n_i;x) =\left \{
    \begin{aligned}
        0,\ & if\ G_i \geq G(n_i;x); \\
        1, \ & otherwise.
    \end{aligned}
    \right.
    \label{eq51}
\end{align}
where $T(n_i;x)$ is an indicator function defined as \eqref{eq51} to indicate whether the data points lie below the curve. $\eta$ denotes the proportion of data segmentation, i.e., the ratio of data points positioned below the fitted curve.
As specified in the proposed program, the first term determines the location of the calibrated curve using a data segmentation parameter, while the second term minimizes the deviation between the fitted curve and the observations based on the MAPE metric. 

A set of MFDs can be derived by adjusting $\eta$ 
at different scales 
to capture the scatter in macroscopic data, as represented in \eqref{eq52}. 
\begin{align}
    \mathcal{G}=\{ G^{(\eta)}(n;x) \ | \ \eta \in \xi \} 
    \label{eq52}
\end{align}
where $\xi$ is a predefined set of $\eta$ (e.g., $\xi = \{0.05, 0.1, \dots, 0.95\}$). $G^{(\eta)}(n;x)$ is the fitted MFD curve obtained by solving the optimization problem \eqref{eq50} with the corresponding $\eta$. 
The ensemble of MFDs captures data scatter without assuming any predefined distributions, accurately reflecting the empirical distribution of observations. 
Furthermore, the approach is flexible as it requires no modifications to the objective function to accommodate varying data distributions.

\subsection{Experimental results}
\label{sec:MFD:cali:case}

\begin{figure}[!htbp]
\centering
\subfigure[Road network of Baoding \label{fig:baoding_road}]{
    {\includegraphics[width=0.24\linewidth]{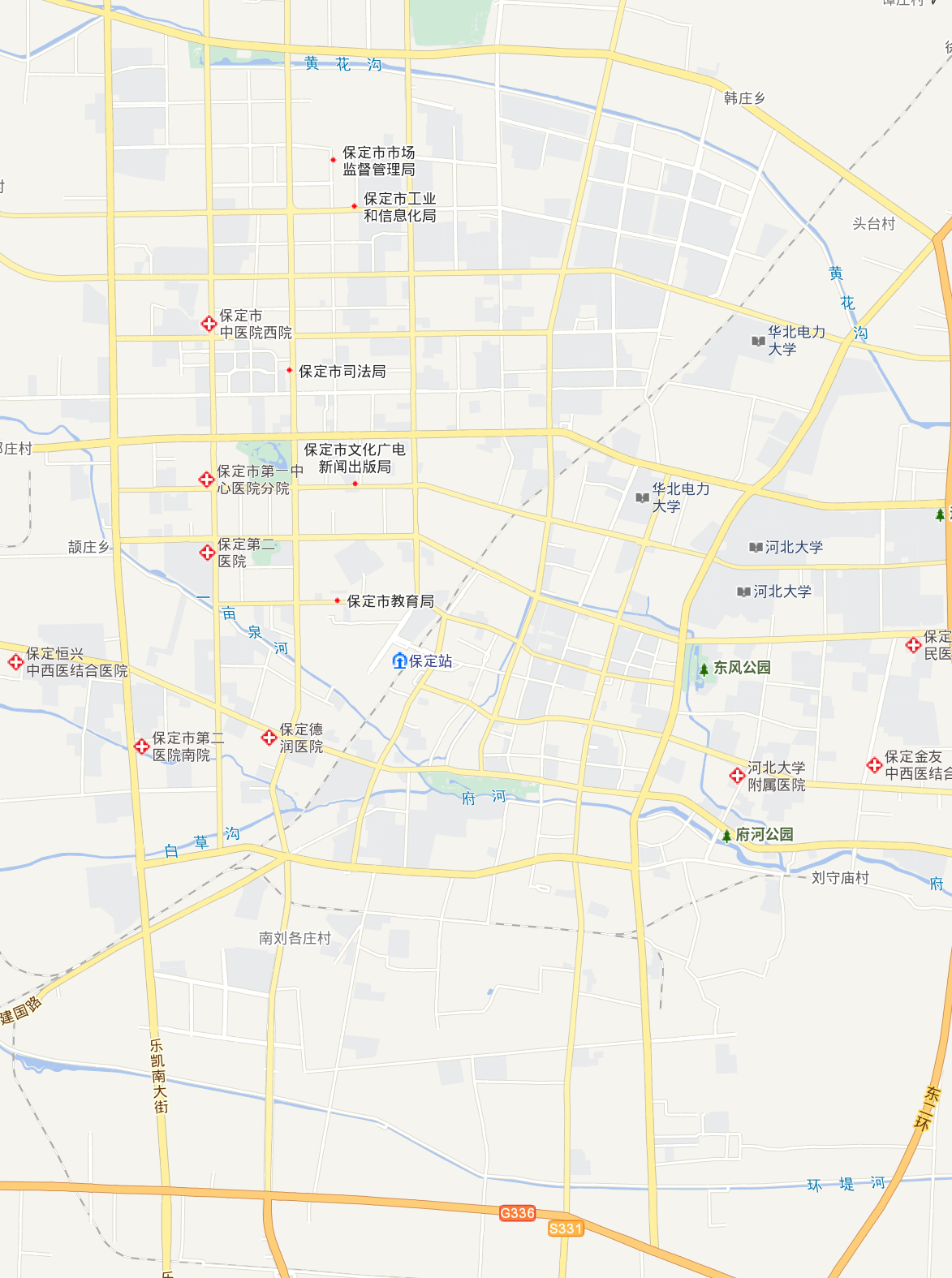}}
}
\subfigure[Topology of Baoding \label{fig:baoding_topo}]{
    {\includegraphics[width=0.24\linewidth]{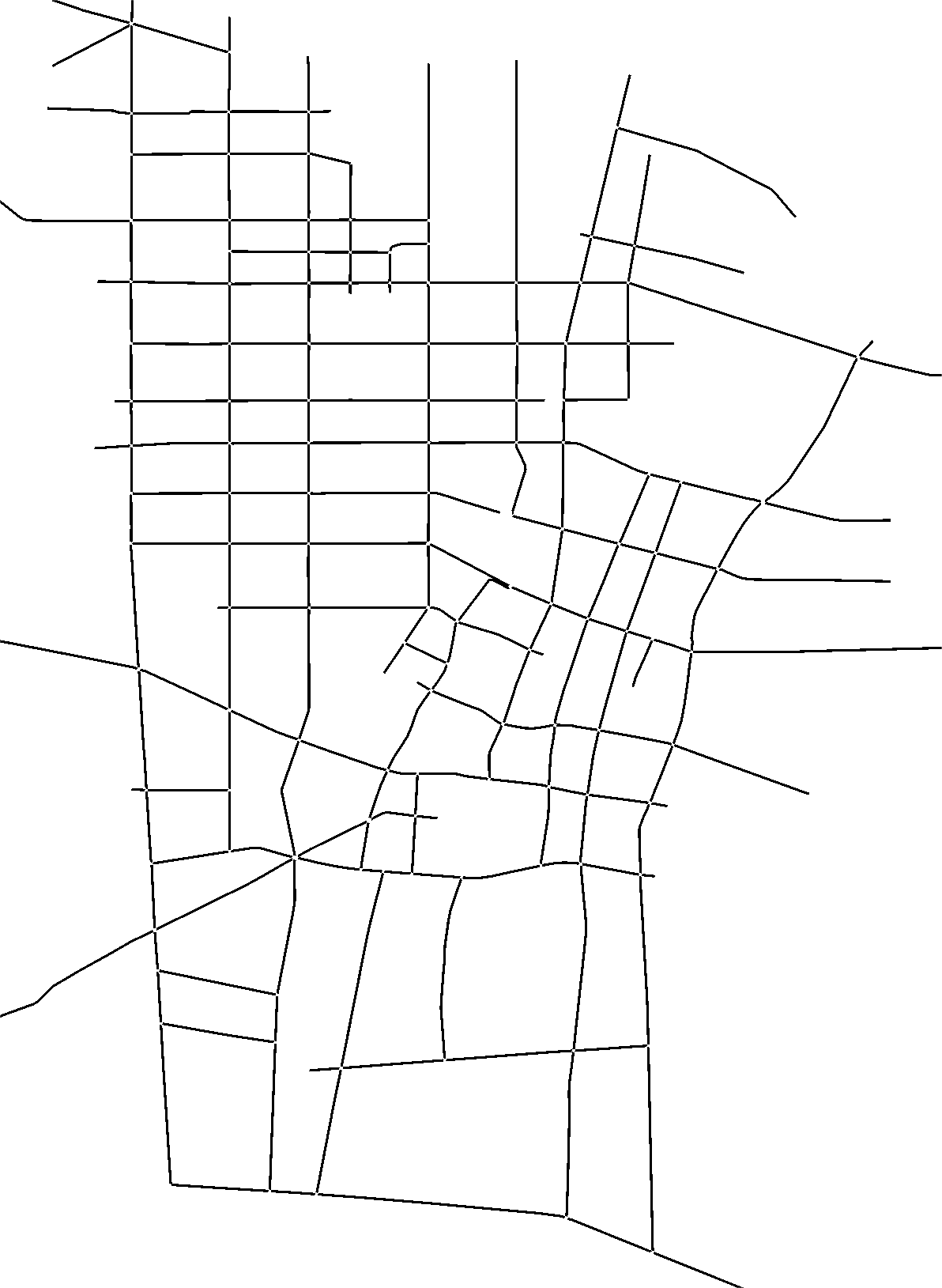}}
}
\subfigure[Macroscopic data of Baoding \kern-0.8cm\label{fig:baoding_data}]{
    {\includegraphics[width=0.45\linewidth]{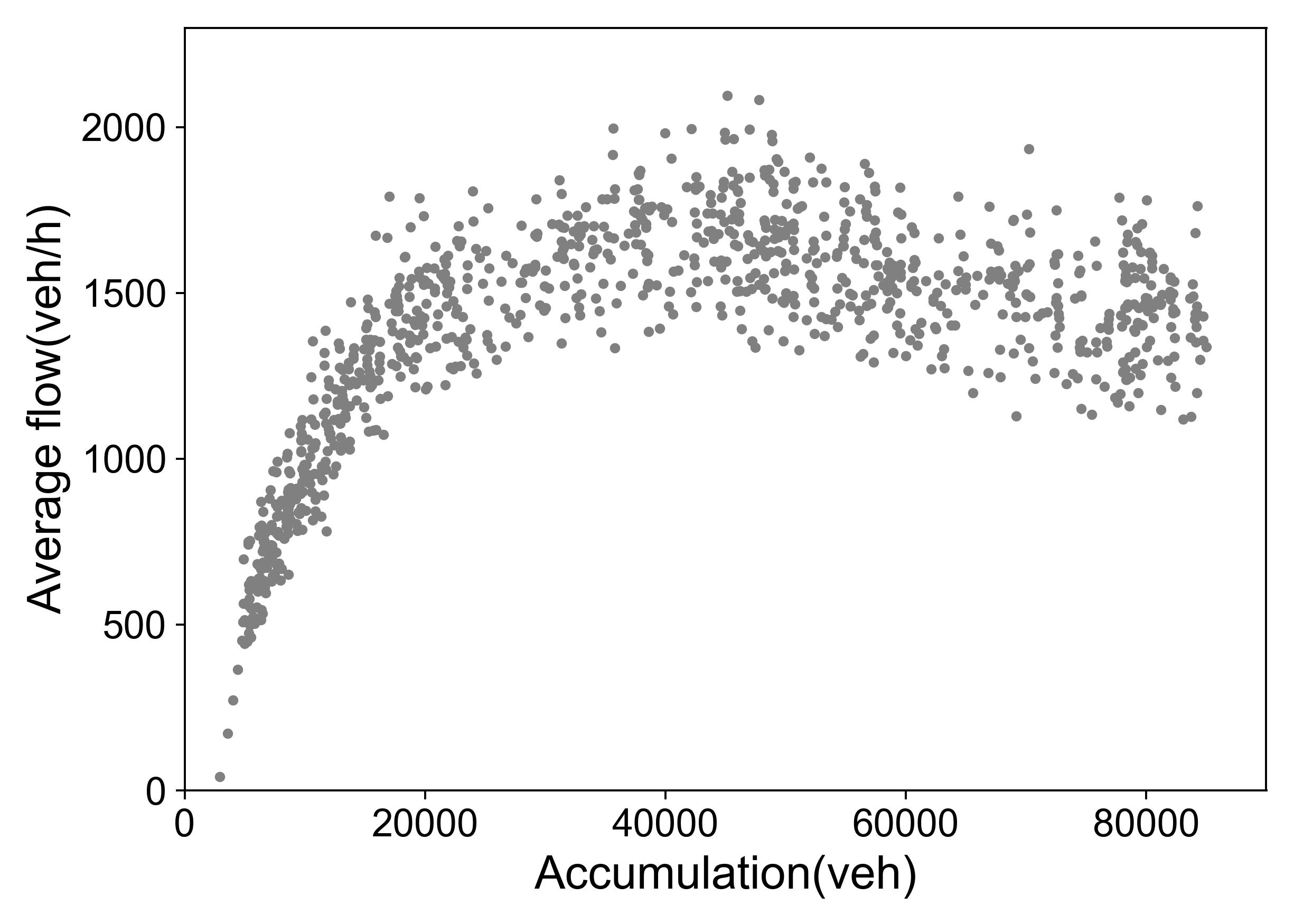}}
}
\subfigure[Road network of Nanchang \label{fig:nanchang_road}]{
    \raisebox{+0.15\height}{\includegraphics[width=0.24\linewidth]{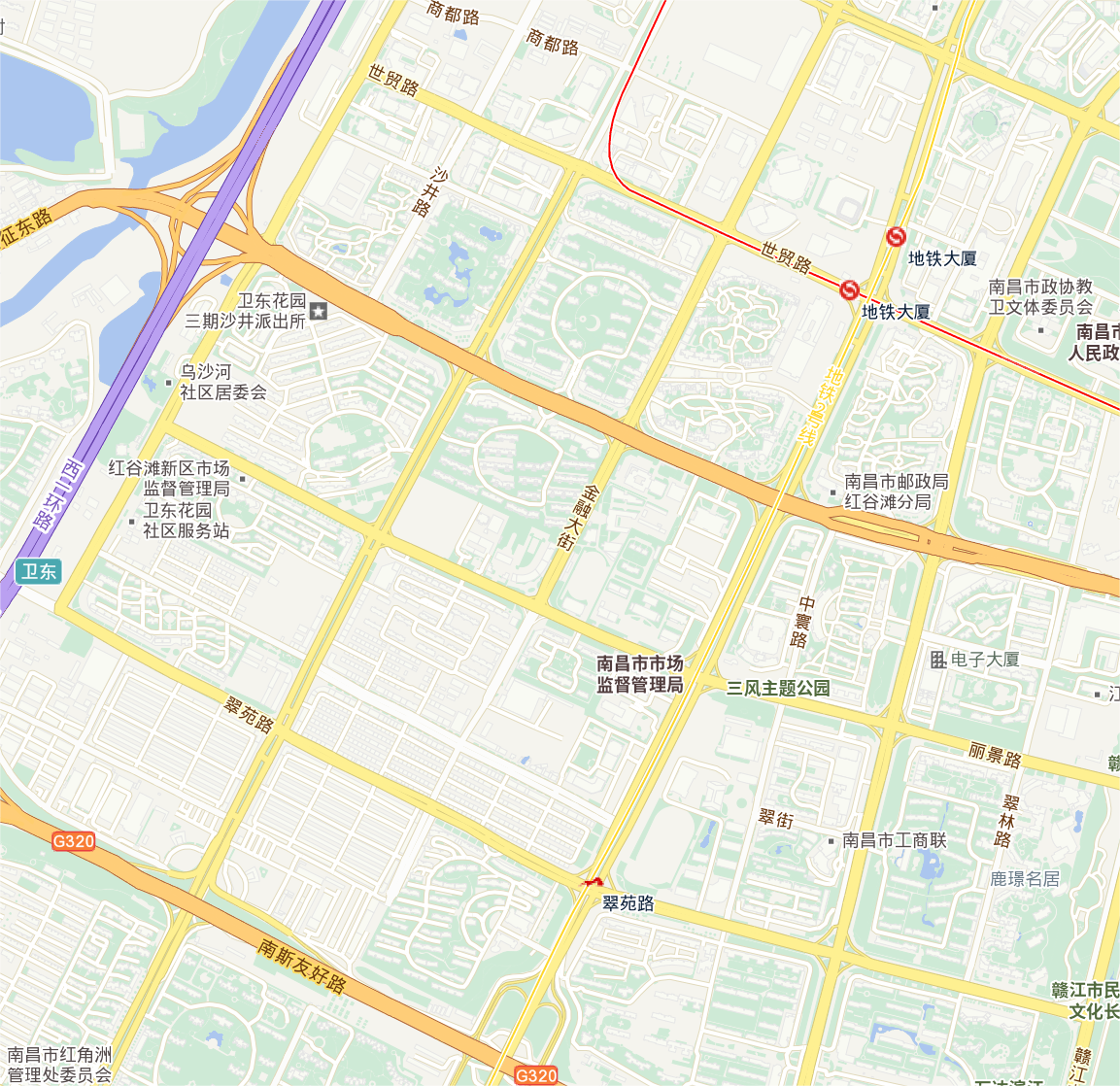}}
}
\subfigure[Topology of Nanchang \label{fig:nanchang_topo}]{
    \raisebox{+0.15\height}{\includegraphics[width=0.24\linewidth]{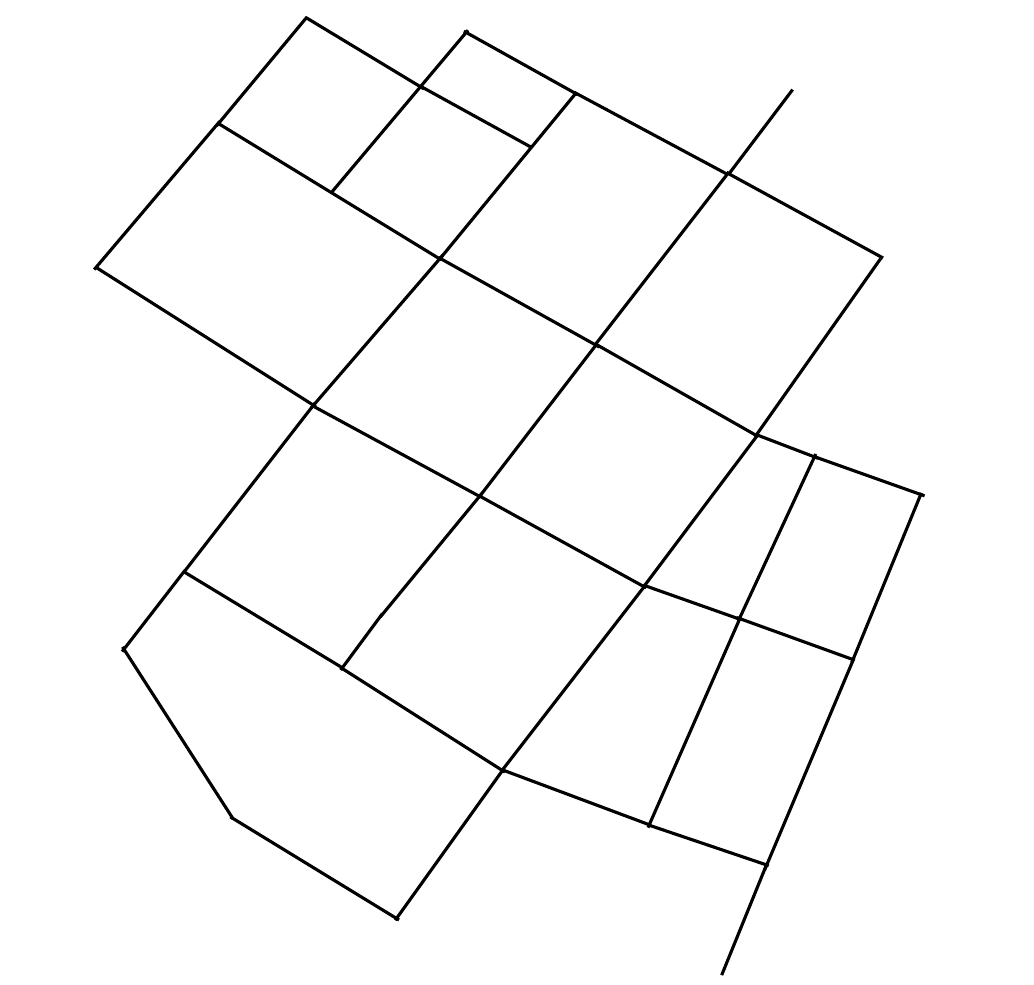}}
}
\subfigure[Macroscopic data of Nanchang \kern-0.8cm\label{fig:nanchang_data}]{
    {\includegraphics[width=0.45\linewidth]{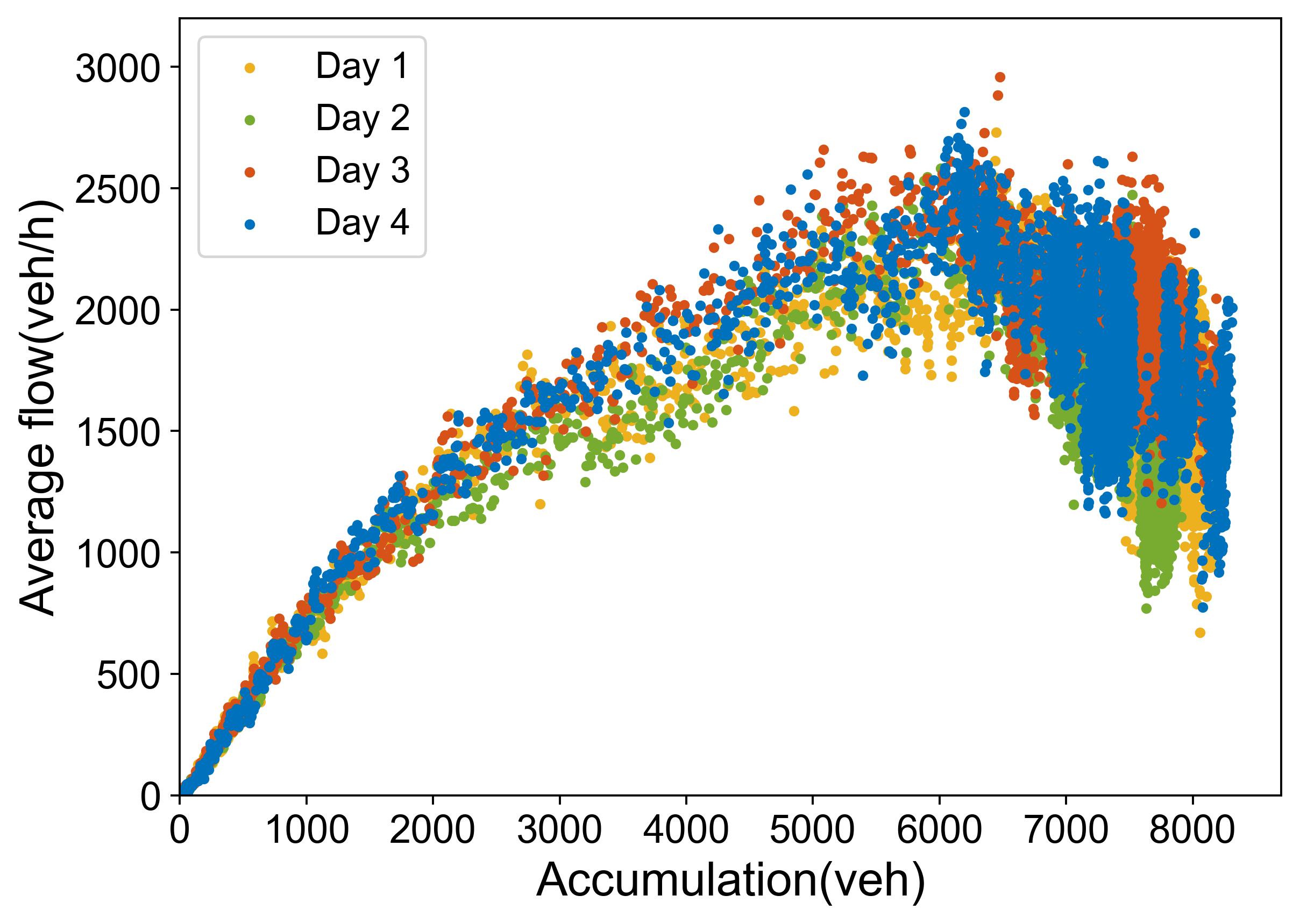}}
}
\caption{Road networks and data of Baoding and Nanchang City.}
\label{fig:road_road}
\end{figure}

Macroscopic data from two urban networks in China are utilized to evaluate the performance of two uncertainty quantification approaches capturing data scatter.
%
The first network is located in Baoding City, Hebei province, as shown in \autoref{fig:baoding_road}-\autoref{fig:baoding_topo}, consisting of 123 intersections and 236 roads. Video surveillance data collected during the peak hour of September 1, 2021, are used to reconstruct origin and destination (O-D) pair demand for simulation, given the sparse distribution of cameras in the network. The simulation lasts for 10,000 seconds, starting from 6:00 am to mimic the morning peak. 
The second network is from Honggutan District, Nanchang City, Jiangxi Province, and comprises 36 intersections and 51 roads, as shown in \autoref{fig:nanchang_road}-\autoref{fig:nanchang_topo}. A total of 200,000 vehicles are distributed over a 10-hour simulation period (8:00-18:00). O-D pairs are randomly selected from intersections within the network, and travel routes are determined based on the underlying traffic assignment model implemented in the simulator. The simulation is conducted four times.

Both networks are simulated using the open-source traffic simulator CityFlow \citep{zhang2019cityflow}, utilizing data collected from these networks. 
Virtual loop detectors are used to gather traffic measurements, including speed $v_i$, density $k_i$, and the accumulation of vehicles $n_i$ in link $i$.
In line with \cite{geroliminis2008existence}, the average flow of a network at time $t$ based on loop detector data is defined as:
\begin{align}
    Q(t)=\frac{\sum_{i=1}^H Q_i(t) l_i}{\sum_{i=1}^H l_i}=\frac{\sum_{i=1}^H (v_i(t) k_i(t)) l_i}{\sum_{i=1}^H l_i}
    \label{q_define}
\end{align}
The accumulation of a network at time $t$ can be calculated as:
\begin{align}
    n(t)=\sum_{i=1}^H n_i(t)
    \label{n_define}
\end{align}
where $H$ denotes the set of equipped links (i.e., links with virtual loop detectors) in the network. $l_i$ is the length of link $i$.

To mitigate data noise, traffic measurements are aggregated at 10-second intervals. The resulting flow and accumulation measurements of Baoding and Nanchang networks are illustrated in \autoref{fig:baoding_data} and \autoref{fig:nanchang_data}, respectively. 
According to \cite{ma2024functional}, the $\lambda$-trapezoidal MFD is chosen for the calibration of Baoding data, while the cubic MFD is chosen for Nanchang data.

\subsubsection{Experiments on incorporating stochasticity into parameters}
\begin{figure}[!htbp]
\centering
\subfigure[Baoding City]{
    \label{baoding_sto_para}
    \includegraphics[width=0.45\linewidth]{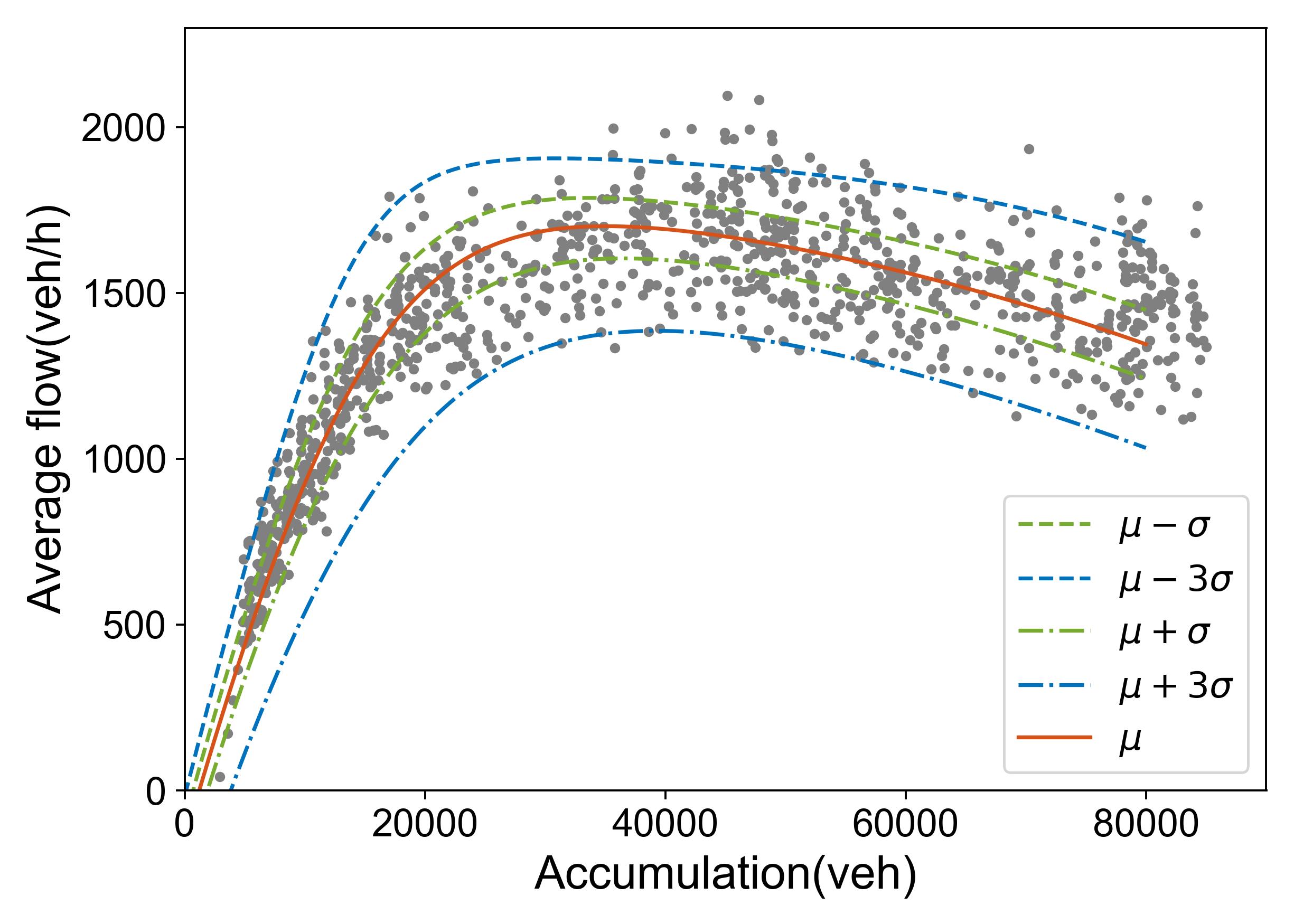}
    }
\subfigure[Nanchang City]{
    \label{nanchang_sto_para}
    \includegraphics[width=0.45\linewidth]{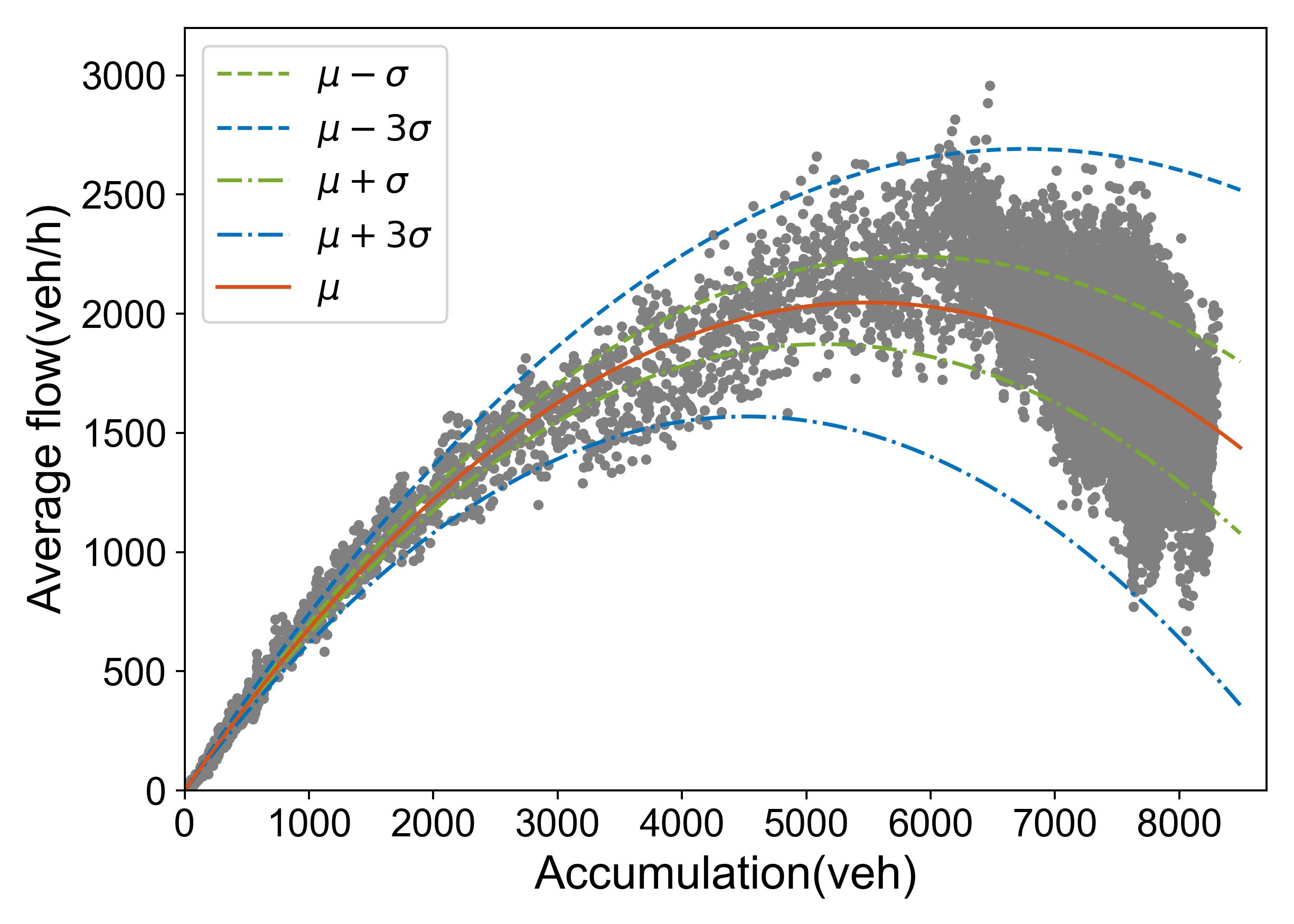}
    }
\caption{Quantification results based on incorporating stochasticity into parameters.}
\label{sto_cali_para}
\end{figure}

As discussed in \autoref{sec:MFD:cali:uncertainty_1}, the smoothing parameter $\lambda$ of the $\lambda$-trapezoidal MFD is selected to incorporate stochasticity. 
Additionally, all parameters in the cubic MFD are assumed to be random variables following Gaussian distributions.
The uncertainty quantification results based on incorporating stochasticity into model parameters are shown in \autoref{baoding_sto_para} and \autoref{nanchang_sto_para}.
The orange solid lines in the figures represent the nominal MFD denoted as $\mu$. The green and blue dashed lines represent the curves for $\mu\pm\sigma$ and $\mu\pm3\sigma$, respectively.
Notably, the extent of uncertainty in MFD varies across different accumulation levels.
In both Baoding and Nanchang, the range of data scatter increases monotonically during the free-flow state, particularly before reaching the critical accumulation of 20,000 $veh$ in Baoding and 6,000 $veh$ in Nanchang. 
In the congestion state, data scatter in Baoding remains relatively stable, whereas in Nanchang, it becomes significantly more pronounced compared to the free-flow state.

\begin{figure}[!htbp]
\centering
\subfigure[Baoding City (40,000 $veh$)]{
    \label{baoding_k40000}
    \includegraphics[width=0.45\linewidth]{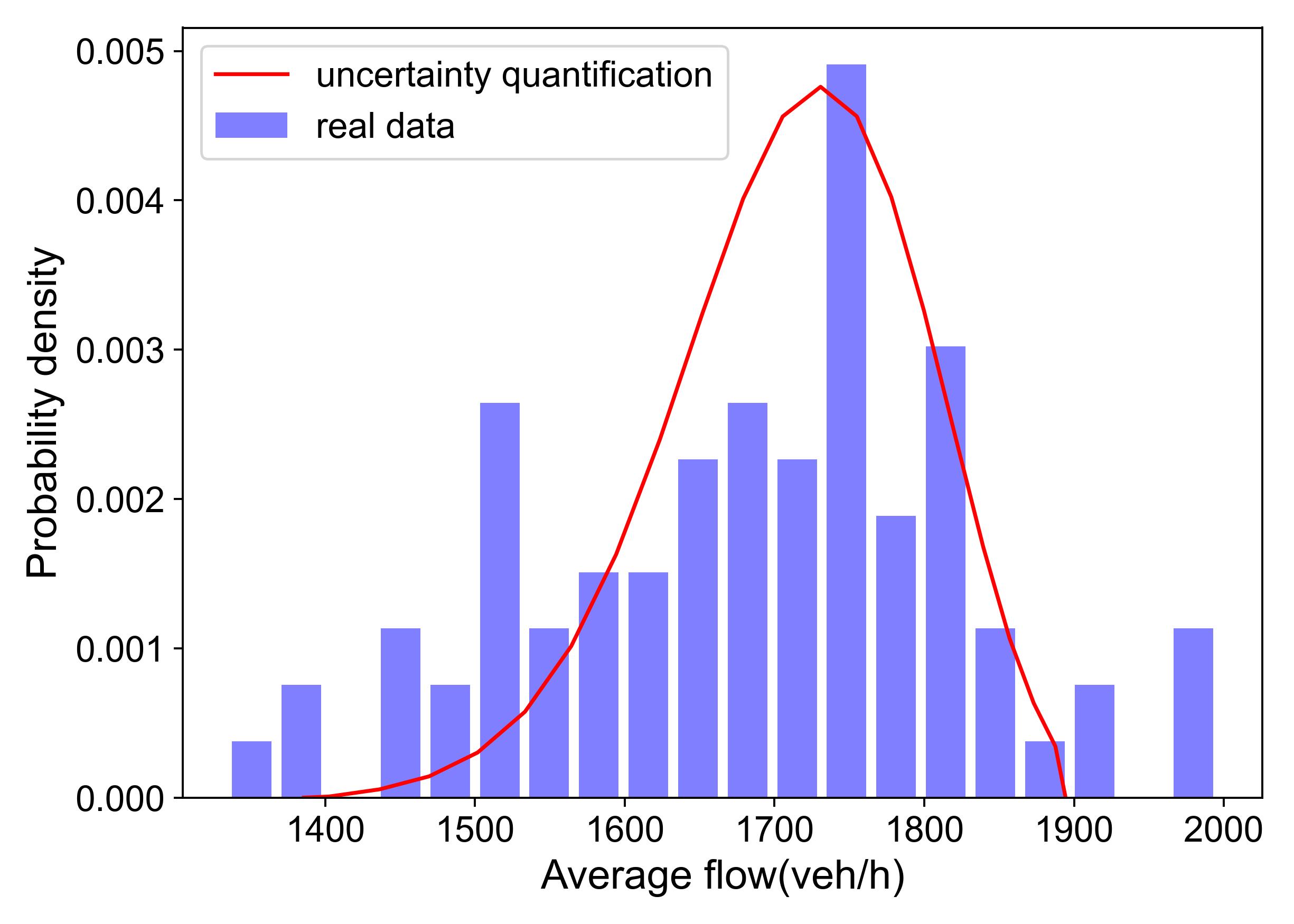}
    }
\subfigure[Nanchang City (7,000 $veh$)]{
    \label{nanchang_k7000}
     \includegraphics[width=0.45\linewidth]{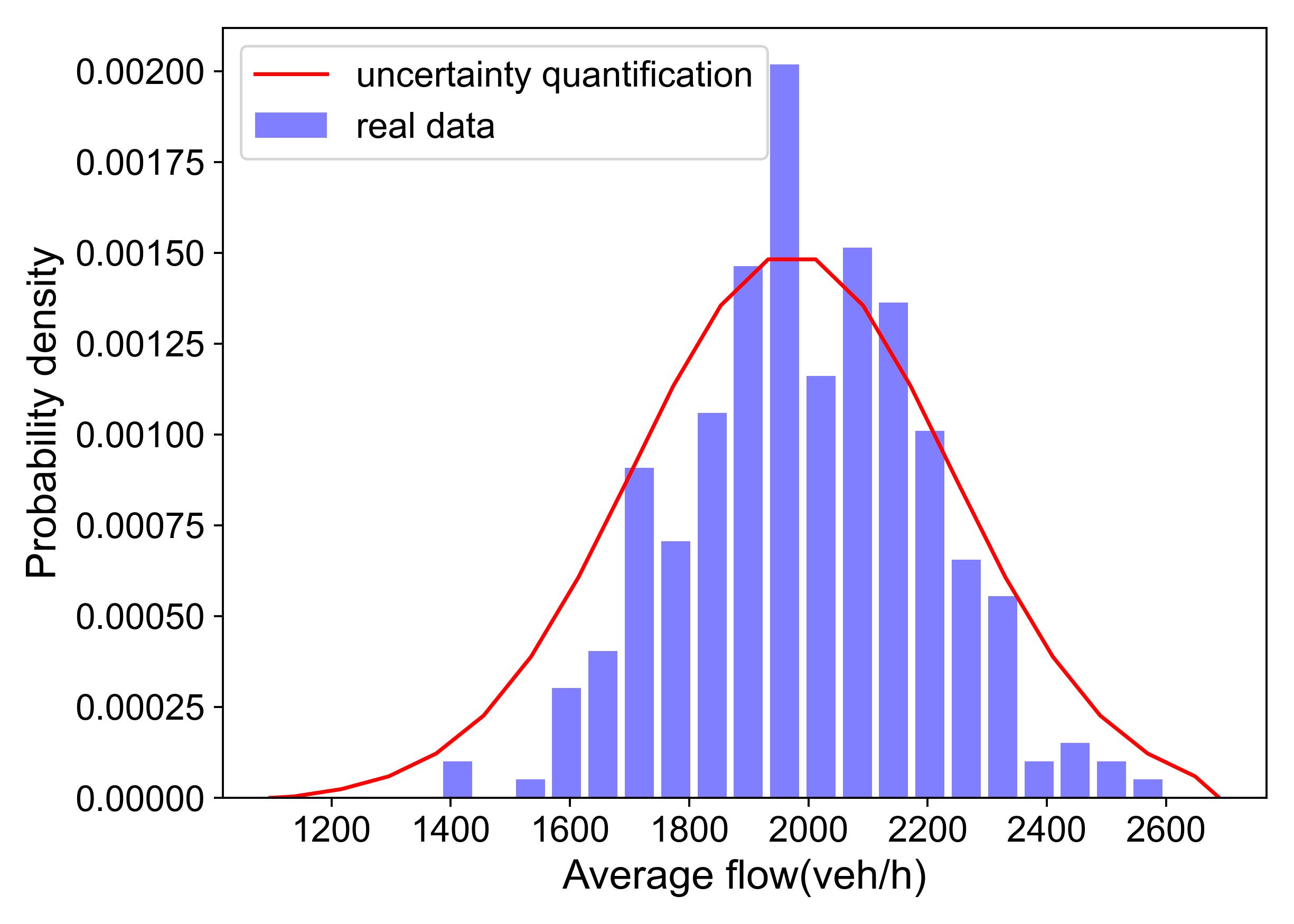}
    }
\caption{Empirical versus quantified probability density at a specific average flow.}
\label{sto_cali_para_probability_density}
\end{figure}

To further evaluate the accuracy of the quantification results in representing the average flow distribution, the probability density functions for Baoding at an accumulation of 40,000 $veh$ and Nanchang at 7,000 $veh$ are presented in \autoref{baoding_k40000} and \autoref{nanchang_k7000}, respectively.
The distribution fitting for Nanchang demonstrates the effectiveness of the quantification approach in capturing the stochastic nature of real data when sufficient data and a well-defined distribution are available. 
However, the results for Baoding highlight the importance of selecting an appropriate predefined distribution, as it influences the accuracy of uncertainty quantification for MFD data, with limited data further amplifying deviations.

\subsubsection{Experiments on calibrating stochastic traffic data}

\begin{figure}[!htbp]
\centering
\subfigure[Baoding City]{
    \label{baoding_sto_distri}
    \includegraphics[width=0.45\linewidth]{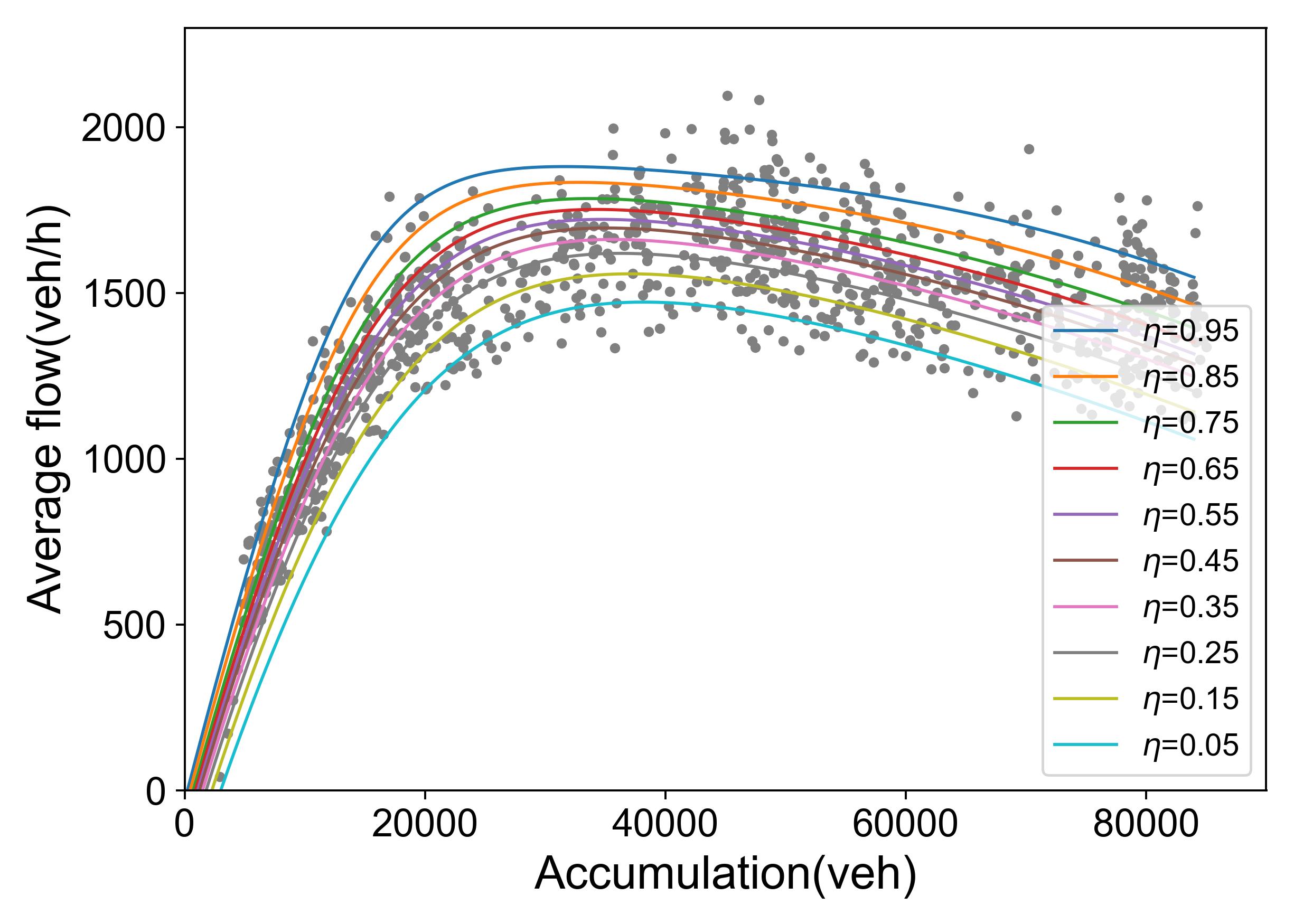}
    }
\subfigure[Nanchang City]{
    \label{nanchang_sto_distri}
    \includegraphics[width=0.45\linewidth]{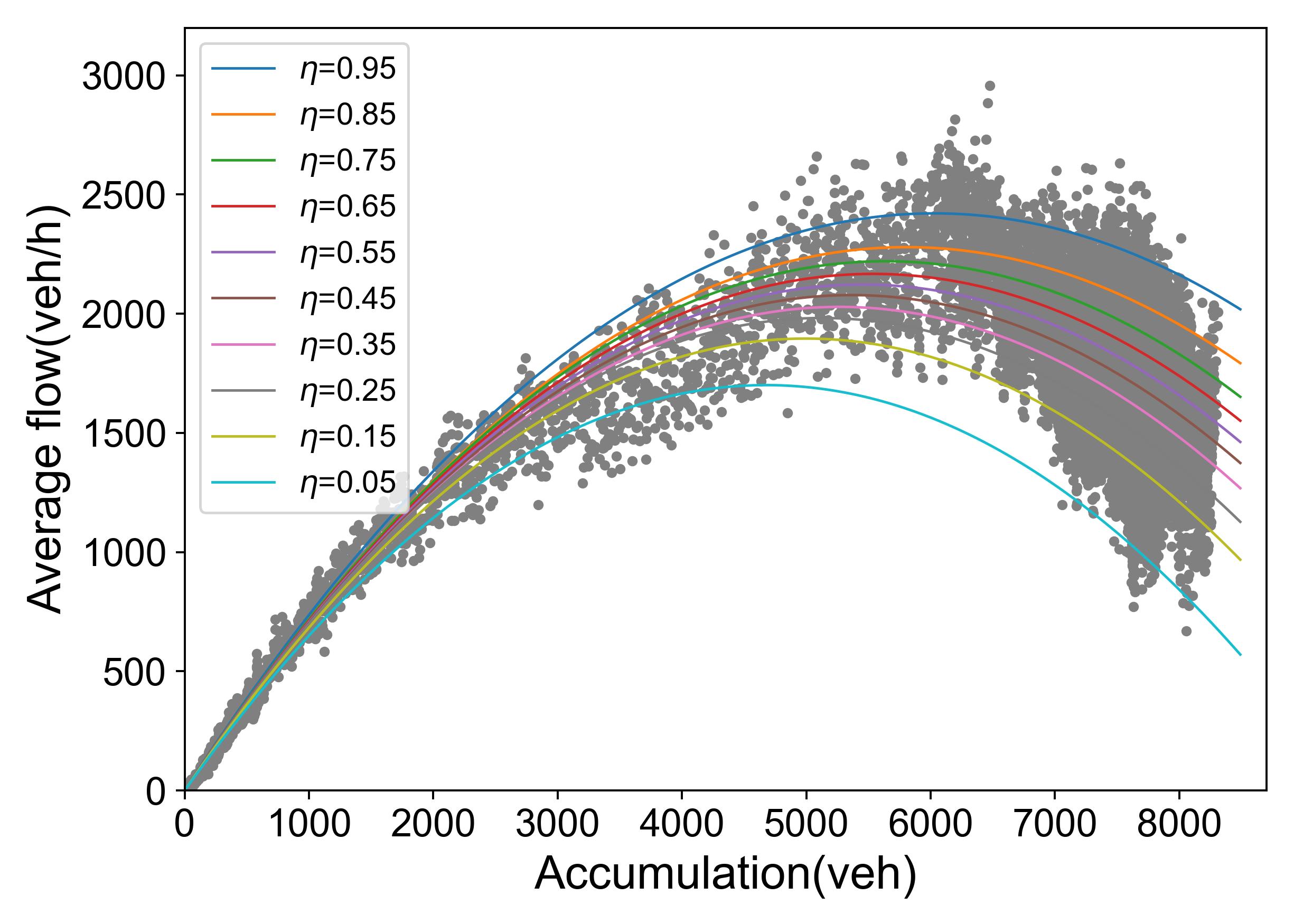}
    }
\caption{Quantification results based on calibrating stochastic traffic data.}
\label{sto_cali_distri}
\end{figure}

The quantification results based on calibrating stochastic traffic data are depicted in \autoref{sto_cali_distri}.
The bottom-up curves represent the calibrated results obtained by incrementally adjusting the data segmentation ratio $\eta$ from small (0.05) to large (0.95). For example, 
for $\eta=0.05$, 5\% of the data lies below the curve, while 95\% lies above it.
Similar to the results shown in \autoref{sto_cali_para}, the uncertainty quantification by calibrating stochastic traffic data effectively captures the scatter of MFD data. Moreover, this approach is more flexible and precise, as the ratio $\eta$ can be arbitrarily set between 0 to 1.

\begin{figure}[!htbp]
    \centering
    \includegraphics[width=0.55\linewidth]{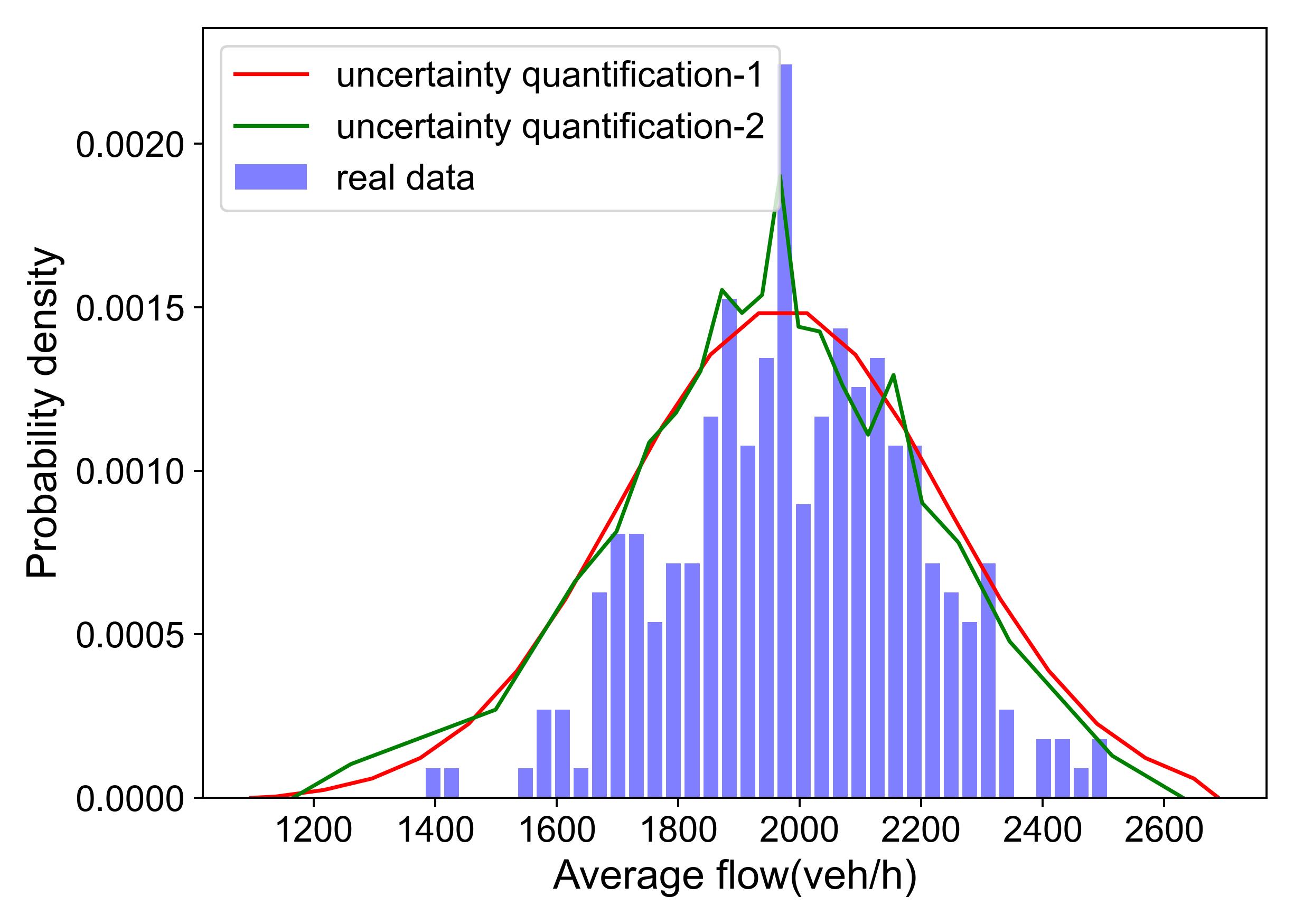}
    \caption{Empirical versus quantified probability density of Nanchang City (7,000 $veh$).}
    \label{nanchang_k7000_compare}
\end{figure}

\autoref{nanchang_k7000_compare} presents a comparison of the probability density of the average flow for Nanchang at a vehicle accumulation of 7,000 $veh$, as determined by two uncertainty quantification approaches. Smaller intervals were employed to achieve a more refined estimation of the empirical data.
The red line represents the result of incorporating stochasticity into parameters, while the green line corresponds to the result of calibrating stochastic traffic data.
%
Both approaches yield similar results for average flow values below 1,800 $veh/h$ and above 2,200 $veh/h$. However, a significant difference emerges for average flow values between 1,800 and 2,200 $veh/h$: the approach assuming a predefined Gaussian distribution fails to capture the abrupt changes observed in this range. 
In contrast, the approach that directly calibrates stochastic data provides a closer approximation to the real flow distribution.
Furthermore, by avoiding prior assumptions about parameters, this approach offers greater flexibility in capturing data scatter, making it more suitable for datasets with varying distribution characteristics.


While these approaches primarily quantify MFD scatter width by measuring data distributions between curves, they face two critical limitations. First, the selection of representative curves to bound MFD uncertainty remains arbitrary. For example, there is no systematic approach to determine appropriate $\sigma$ levels for uncertainty representation when incorporating stochasticity into model parameters. Second, the selected uncertainty boundaries lack physical interpretation, hindering their practical application in traffic control and management strategies.
\section{Uncertainty quantification approach considering distinct congestion phases}\label{sec:MFD:cali:quantification}



Prior research has shown that congested traffic networks exhibit distinct loading and recovery phases, known as the hysteresis phenomenon, leading to MFDs characterized with significant scatter \citep{geroliminis2011Hysteresis, xu2023opposing}.
This observation suggests that an uncertainty quantification approach that explicitly accounts for the different physics of congestion loading and recovery may offer greater physical interpretability.
Therefore, this section proposes a novel approach that considers distinct congestion phases and evaluates its effectiveness using microscopic simulation data. 


\subsection{MFD hysteresis and segmentation method}
In this subsection, we first construct a microscopic simulation to generate macroscopic data.
Subsequently, we utilize this simulated data to illustrate the MFD hysteresis phenomenon and present a segmentation method that divides MFD hysteresis into loading and recovery phases.

\begin{figure}[!htbp]
\centering
\subfigure[The simulated grid network \kern 0.9cm \kern1.2cm\label{fig:topo_pc}]{
    {\includegraphics[width=0.45\linewidth]{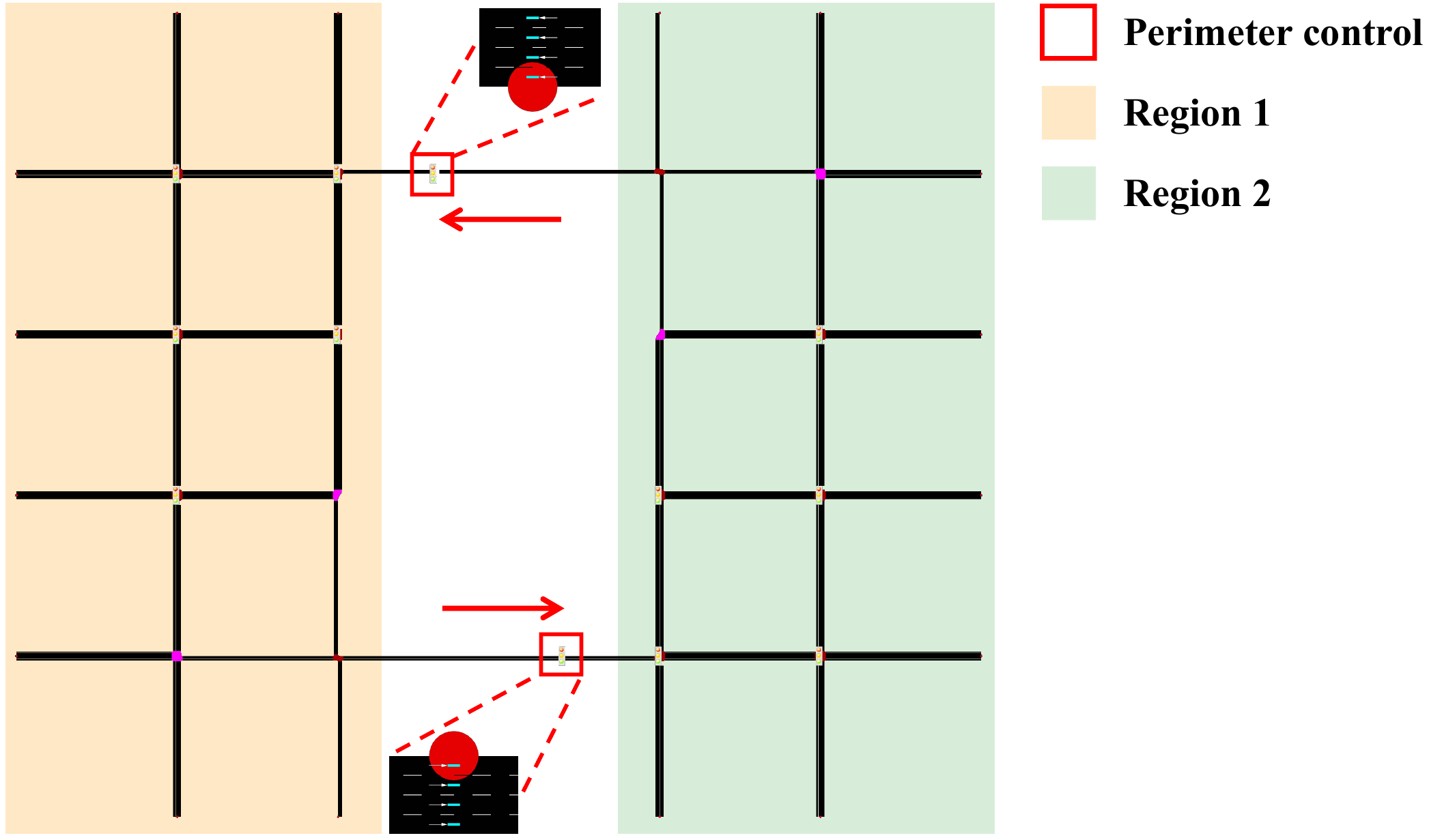}}
}
\subfigure[The aggregated data \kern 0.8cm\label{fig:region2_color}]{
    {\includegraphics[width=0.45\linewidth]{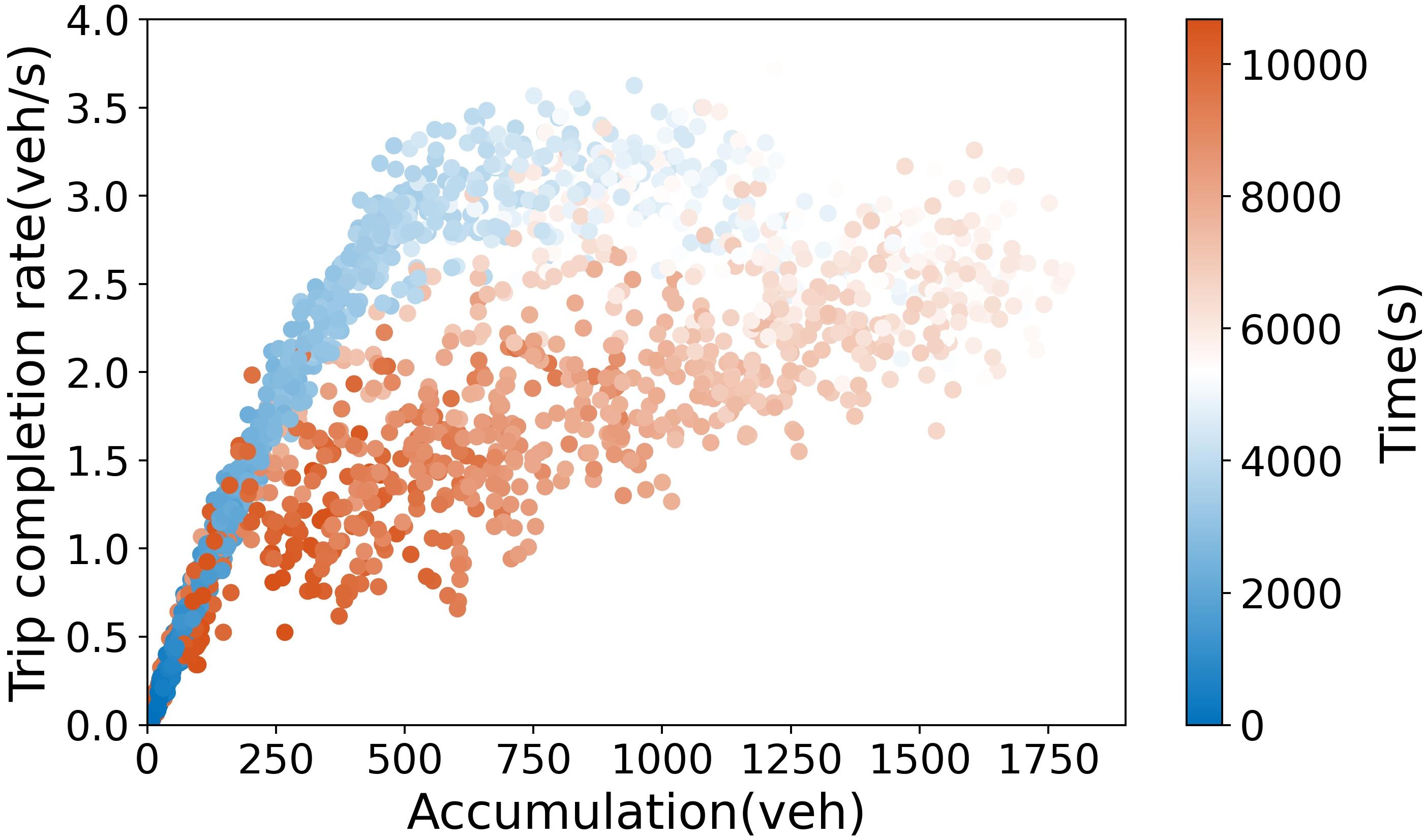}}
}
\caption{The topology and aggregated data of the microscopic simulation.}
\label{fig:simu_pc}
\end{figure}

The simulated grid network in \autoref{fig:topo_pc} comprises two regions colored orange and green, respectively. The network includes 70 roads and 18 intersections.
Within each region, roads are 500 meters in length with 4 lanes.
Boundary roads connecting adjacent regions span 1,000 meters, with 700 meters extending upstream of the perimeter control intersections.
The simulation runs for a total of 3 hours and is implemented using the open-source traffic simulator CityFlow \citep{zhang2019cityflow}.

Using Region 2 as a case study, we collect data from 20 simulation replications, aggregated at 120-second intervals. \autoref{fig:region2_color} visualizes the aggregated accumulation and trip completion rate data, distinguishing between congestion loading (cool colors) and recovery (warm colors) processes.
During the free-flow state (network accumulation $<$ 750 $veh$), the data exhibits low scatter with the trip completion rate monotonically increasing until reaching its maximum of approximately 3.5 $veh/s$ at 750 $veh$ accumulation.
Beyond this critical threshold of 750 $veh$, the network enters a congested state characterized by declining performance. The trip completion rate deteriorates significantly, dropping to 70\% of its peak value when network accumulation reaches its maximum of approximately 1750 $veh$.
The data demonstrate a pronounced clockwise hysteresis pattern: trip completion rates are systematically higher during congestion loading (increasing accumulation) compared to congestion recovery (decreasing accumulation). 
This pattern indicates a significant capacity drop during network recovery from severe congestion \citep{li2023perimeter}.





\begin{figure}[!htbp]
\centering
\subfigure[Calibration of full aggregated data \kern -0.5cm\label{fig:full_cali}]{
    {\includegraphics[width=0.45\linewidth]{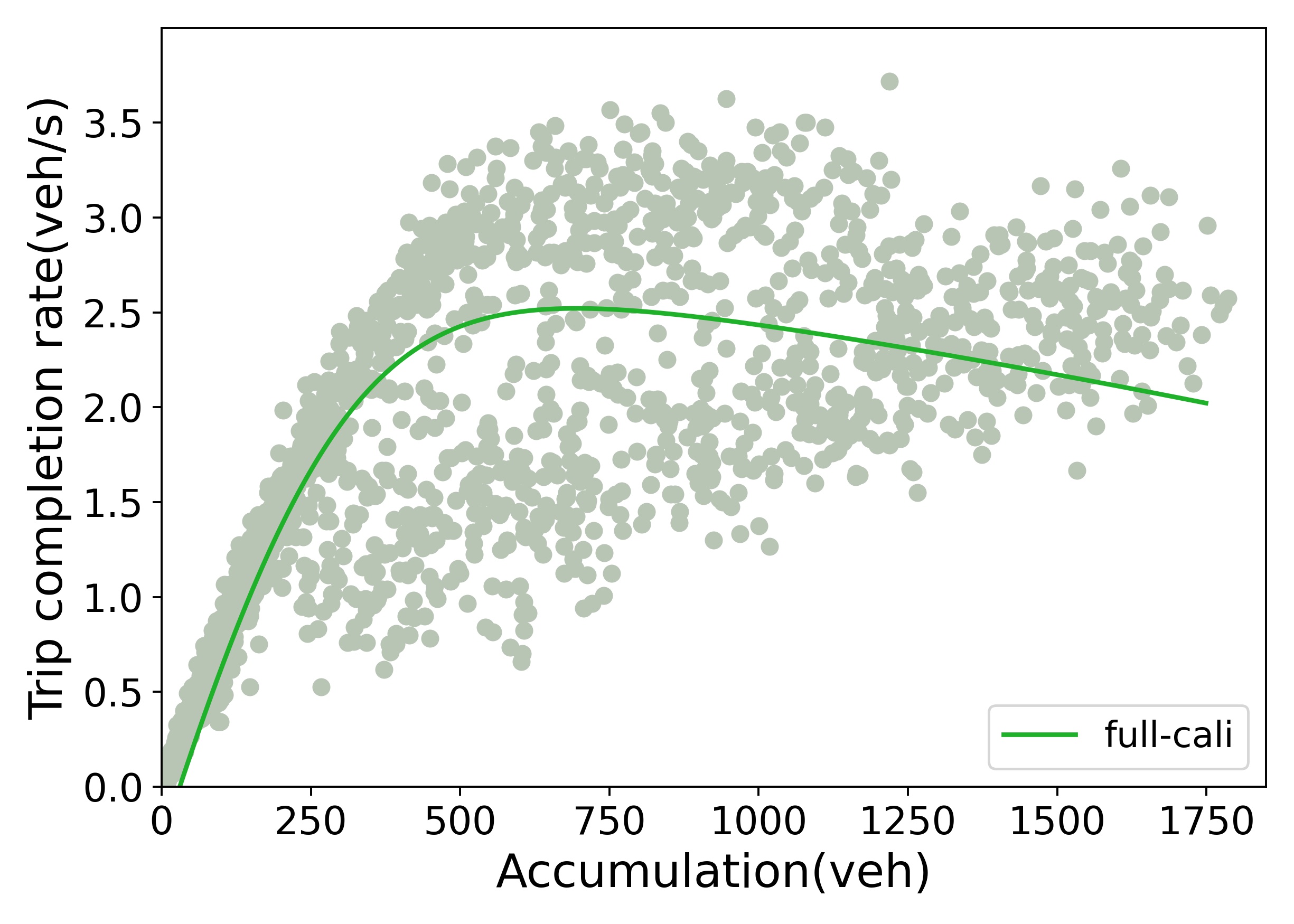}}
}
\subfigure[Calibration of loading and recovery phases \kern -0.5cm\label{fig:loading_cali}]{
    {\includegraphics[width=0.45\linewidth]{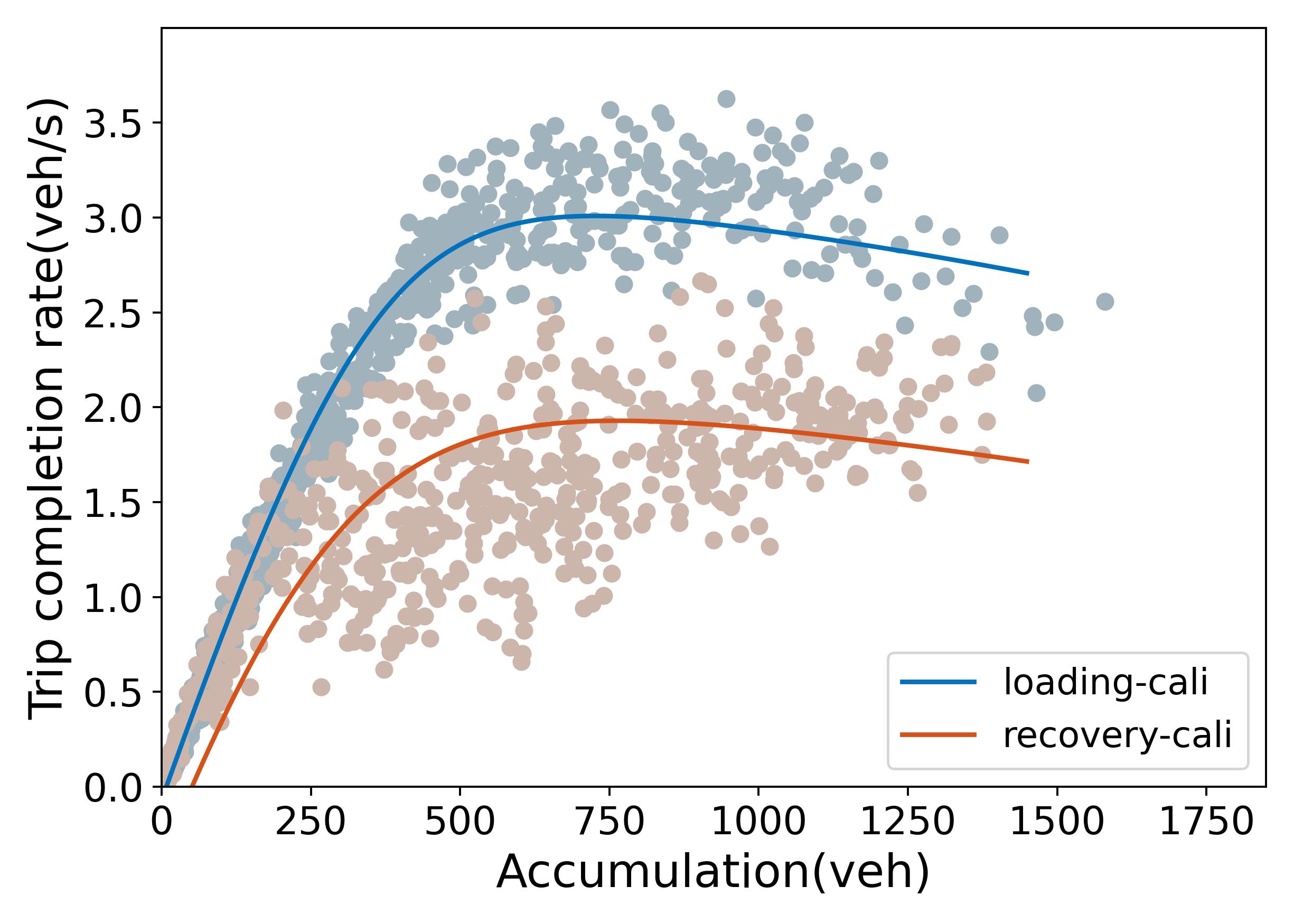}}
}
\caption{Comparison of MFD calibration from full data against loading and recovery phases.} 
\label{fig:simu_cali}
\end{figure}

Prior research has demonstrated that curves calibrated passing through the center of data distributions do not necessarily yield optimal calibration results \citep{zhong2016automatic}. 
This limitation becomes particularly significant when data exhibits notable scatter and uncertainty, as shown in \autoref{fig:region2_color}.
Direct calibration of such MFD data using the $\lambda$-trapezoidal function \citep{ambuhl2020afunctional}, as shown in \autoref{fig:full_cali}, exhibits systematic biases. While the fitted curve accurately captures the free-flow state, it bisects the hysteresis loop during congestion, leading to two critical estimation errors: underestimating trip completion rates during congestion loading and overestimating during recovery. Consequently, the calibration fails to represent the underlying traffic dynamics faithfully.

These findings point to the necessity of calibrating congestion loading and recovery phases independently. While several methods have been proposed to distinguish between these phases \citep{paipuri2019validation, leclercq2019macroscopic, mousavizadeh2024important}, we extend the approach of \cite{mousavizadeh2024important} to develop a three-phase segmentation method. 
The method partitions macroscopic data into: 
congestion loading phase, transitional phase without distinct loading or recovery characteristics, and congestion recovery phase. 
The temporal boundaries between these phases are defined as follows:
\begin{subequations}
\begin{align}
    t_{l,tra} &=  t,\; if \; n(t)<n(t-\alpha \times T_{agg})
    \label{eq:data_seg_ld}
    \\
    t_{tra,re} &= t,\; if \; n(t)<n(t-\beta \times T_{agg})
    \label{eq:data_seg_unld}
\end{align}
\end{subequations}
where $T_{agg}=120s$ denotes the interval of data aggregation. $t_{l,tra}$ is the temporal boundary from the loading phase to the transitional, and $t_{tra,re}$ is the boundary from the transitional to the recovery phase.
The coefficients $\alpha$ and $\beta$ are time horizon parameters used to calculate these temporal boundaries. In our study, the simulation duration is 3 hours, with data aggregated in 120-second intervals, resulting in a total of 90 time steps ($t \leq 90$). Considering data noise and dispersion, the parameters $\alpha$ and $\beta$ are set to 3 and 16, respectively.

\autoref{fig:loading_cali} presents the calibration results for congestion loading and recovery phases, respectively.
This dual-MFD approach achieves superior accuracy in capturing traffic dynamics compared to the single-MFD calibration shown in \autoref{fig:full_cali}.
This bifurcated representation has direct practical applications in traffic control and management. The loading MFD effectively characterizes pre-peak periods when network traffic remains unsaturated, while the recovery MFD captures the capacity drop during mid- and post-peak periods following severe congestion. The integration of both MFDs enables more precise estimation of traffic dynamics, potentially improving the effectiveness of MFD-based control strategies.

\subsection{Uncertainty quantification considering distinct congestion phases}
The MFD uncertainty quantification relies on defining physically interpretable boundary curves to guide effective traffic control and management strategies.  
Although the dynamics of MFD with hysteresis can be effectively captured by segmenting the loading and recovery phases, distinguishing between these two phases remains challenging, especially when MFD data lacks time stamps. Moreover, calibrating two separate MFDs introduces additional complexity compared to calibrating a single curve. 
Given these challenges, incorporating the distinct phases of congestion loading and recovery into the quantification process may provide a more accurate and comprehensive representation of MFD data scatter.

Based on the proposed mathematical program in \eqref{eq:general_quantify} and the conventional quantification principle introduced in \autoref{sec:MFD:cali:uncertainty_1}, the upper bound parameters $x_{up}$ and the distributional parameter of $\lambda$ (i.e., $\mu_\lambda$ and $\sigma_\lambda$) can be calibrated.
Subsequently, we propose an approach that adjusts the coefficient $\delta$ of $\sigma_\lambda$ to approximate distinct congestion phases. The objective function is defined as follows:
\begin{align}
    \min_\delta H(\delta)=
    \frac{1}{M_{ld}} \sum^{M_{ld}}_{i=1}
    \left|\frac{G_{u,\delta}(n_i)-G_i} {G_i}\right|+
    \frac{1}{M_{re}} \sum^{M_{re}}_{i=1}
    \left|\frac{G_{l,\delta}(n_i)-G_i} {G_i}\right|
    \label{eq:ld_unld_cali}
\end{align}
subject to
\begin{subequations}
\begin{align}
    G_{l,\delta} (n) =- \lambda_{u,\delta}\ln (e^{-\frac{k_l n}{\lambda_{u,\delta}}} & +e^{-\frac{G_{max}}{\lambda_{u,\delta}}}+e^{-\frac{k_r(n_{jam}-n)}{\lambda_{u,\delta}}})
    \\
    G_{u,\delta} (n) =- \lambda_{l,\delta}\ln (e^{-\frac{k_l n}{\lambda_{l,\delta}}}& +e^{-\frac{G_{max}}{\lambda_{l,\delta}}}+e^{-\frac{k_r(n_{jam}-n)}{\lambda_{l,\delta}}})
    \\
     \lambda_{u,\delta} =\mu_{\lambda}+\delta  \sigma_{\lambda}, & \quad 
    \lambda_{l,\delta}=\mu_{\lambda}-\delta \sigma_{\lambda}
    \\
    \lambda \sim N & (\mu_{\lambda},\sigma_{\lambda})
\end{align}
\end{subequations}
where $G_{u,\delta}(n), G_{l,\delta}(n)$ denote the upper and lower bound curves defined by parameters $\lambda_{l,\delta}, \lambda_{u,\delta}$, respectively. $D^{M_{ld}},D^{M_{re}}$ are the datasets of congestion loading and recovery, and $M_{ld}, M_{re}$ denote the size of datasets.
The deviation between the quantified curves and the observed data is evaluated using the MAPE metric.
These two terms jointly determine the position of the quantified curves, thereby enabling simultaneous representation of both congestion loading and recovery phases.

\begin{figure}[ht]
    \centering
    \includegraphics[width=0.55\linewidth]{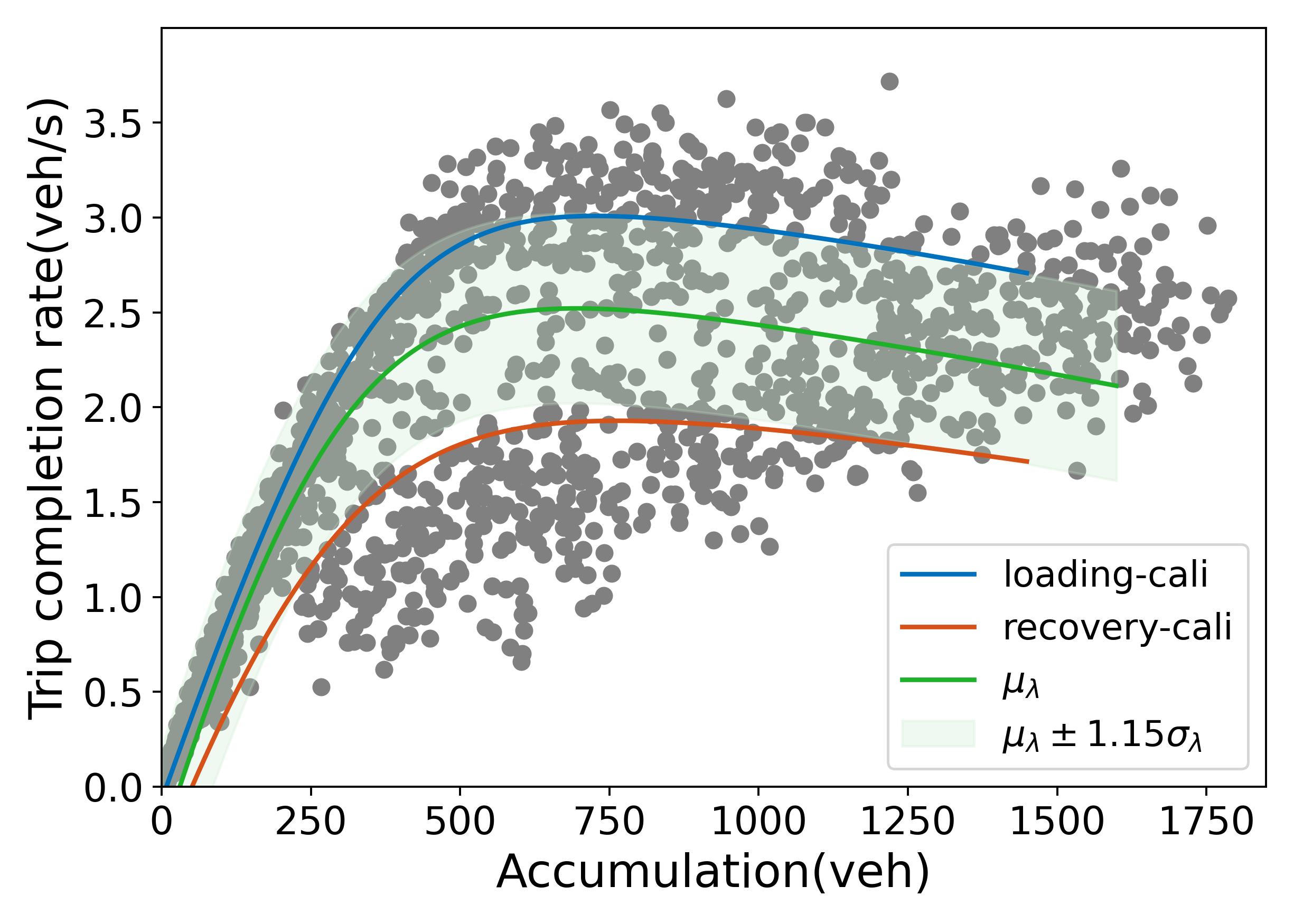}
    \caption{MFD uncertainty quantification against loading and recovery calibration.}
    \label{fig:final_cali}
\end{figure}

The uncertainty quantification results, using the simulated data presented in \autoref{fig:region2_color}, are depicted as the green line with light green shading in \autoref{fig:final_cali}.
Besides, we calibrate the congestion loading and recovery phases separately by employing the segmentation method described in \eqref{eq:data_seg_ld}-\eqref{eq:data_seg_unld}, depicted as blue and orange solid lines, respectively.
The upper bound of the quantified interval aligns with the calibrated congestion loading phase, while the lower bound approximates the recovery phase. This validates that the proposed uncertainty quantification approach presented in \eqref{eq:ld_unld_cali} effectively captures the physics of congestion loading and recovery phases.
Additionally, the calibrated coefficient $\delta$ can be interpreted as an indicator of the distance between network loading and recovery, indicating the capacity drop caused by severe congestion.
The estimated capacity drop, identified by $\delta$, is approximately 15\%, representing the reduction from the average maximal trip completion rate in the free-flow state to the recovery state.
Previous studies have reported that the capacity drop of highways generally lies in the range of 10-20\% \citep{srivastava2013empirical}, supporting the validity of the quantification denoted by $\delta$.

For networks exhibiting pronounced hysteresis in MFDs, model-based control strategies based on deterministic calibration may become suboptimal or even impractical, especially during the network recovery phase. The proposed quantification approach provides a lightweight way to capture the loading and recovery phases of congestion, improving the representation of traffic dynamics. Traffic control strategies tailored to distinct network phases would be better suited for managing real-world traffic conditions.
Moreover, traffic control design can incorporate the parameter $\delta$ into objective functions, such as minimizing total travel time or maximizing network throughput, or use it as a regularization term. 
For example, the MFD function with parameter $\delta$ can be integrated into the objective of maximizing network throughput, thereby improving traffic efficiency while reducing MFD uncertainty.

\section{Influencing factors and traffic resilience analysis} \label{sec:MFD:cali:influence_resilience}


The uncertainty quantification approach proposed in \autoref{sec:MFD:cali:quantification} establishes a connection between the hysteresis phenomenon and MFD uncertainty, providing a theoretical basis for analyzing factors influencing uncertainty.
This section investigates the effects of macroscopic factors, including travel demand and traffic control, and microscopic factors, including vehicle acceleration and deceleration, on network capacity and uncertainty using the MFD hysteresis representation. 
Additionally, a traffic resilience analysis is conducted based on these factors related to MFD uncertainty.

\subsection{Influencing factors analysis}

In this subsection, we explore the factors influencing the hysteresis phenomenon in MFDs, which further affect network capacity and uncertainty.
Given the considerable impact of travel demand, traffic control, vehicle acceleration and deceleration on the shape of MFDs, Region 2 in the simulated network (\autoref{fig:topo_pc}) is selected as an example to examine their effects.
The detailed settings for each scenario are provided in \autoref{appendix_setting}.

\subsubsection{Travel demand}

\begin{figure}[!htbp]
\centering
\includegraphics[width=0.45\linewidth]{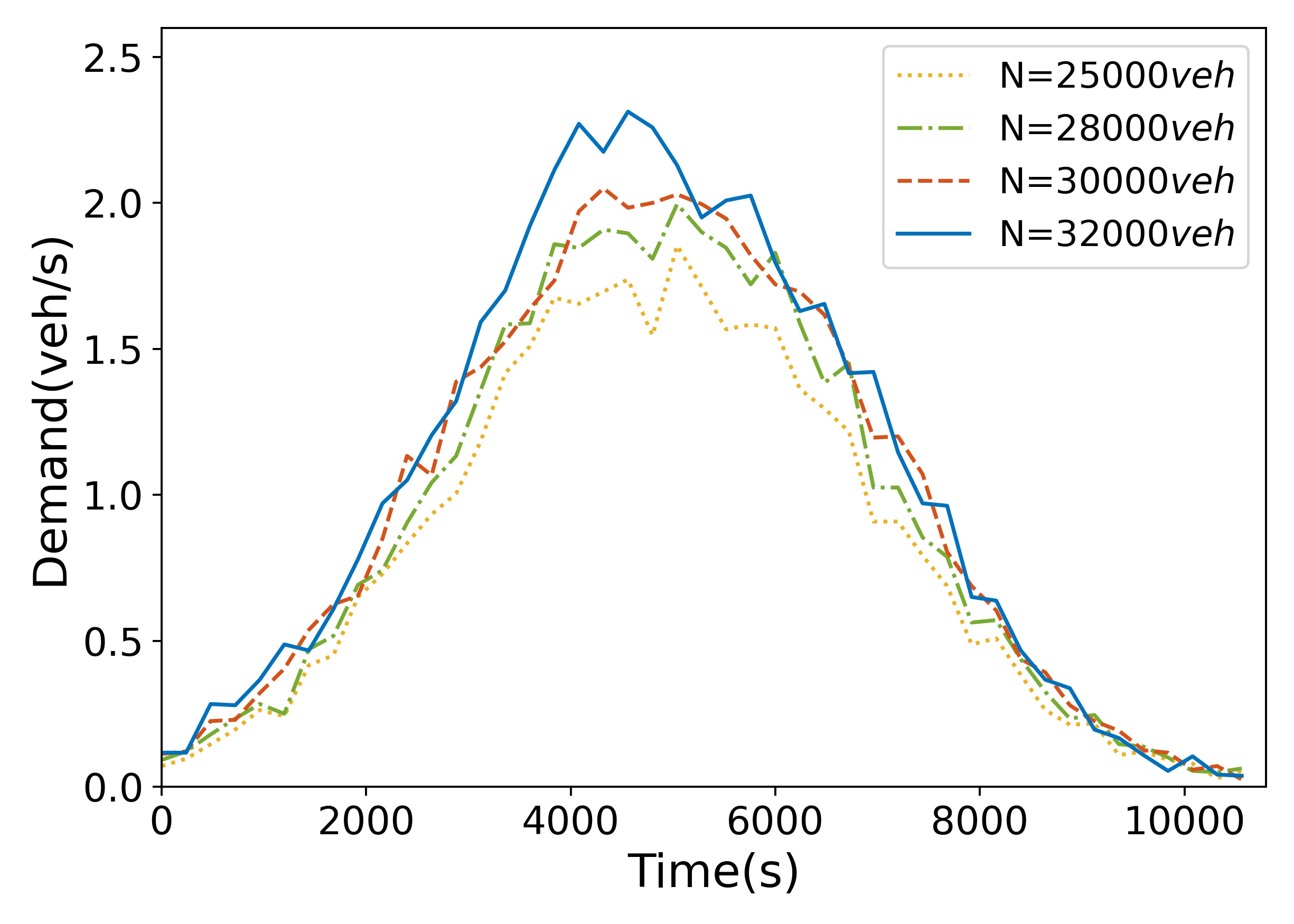}
\caption{The diverse time-varying demand patterns.}
\label{fig:demand_patterns}
\end{figure}

To investigate the relationship between travel demand and MFD hysteresis, we compare MFDs under various time-varying demand patterns. These patterns emulate the peak-hour traffic, including congestion loading and recovery processes.
Total demand inputs denoted as $N$, are set at 25,000, 28,000, 30,000, and 32,000 $veh$. Time-series demand is generated from a normal distribution based on these total inputs, as depicted in \autoref{fig:demand_patterns}, with the region initially empty.

MFDs aggregated at 10-minute intervals under different demand patterns are depicted in \autoref{fig:MFD_demands}. A discernible clockwise hysteresis is observed in all cases, but with different sizes. 
Under lower demand inputs of 25,000 and 28,000 $veh$, smaller hysteresis loops are observed, with the maximal accumulations of 750 and 1000 $veh$, respectively. 
However, MFDs under higher demand inputs of 30,000 and 32,000 $veh$ exhibit notably larger hysteresis loops. The maximal accumulations are 1600 $veh$ and 1700 $veh$, respectively, significantly exceeding the values observed under lower demand inputs. 

These findings emphasize the impact of travel demand on MFD hysteresis, aligning with previous research \cite{leclercq2019macroscopic, xu2023opposing, lu2024traffic}.
Since the hysteresis phenomenon significantly contributes to the uncertainty of MFDs, larger hysteresis loops are associated with greater uncertainty. 
Consequently, traffic demand management (TDM) strategies, such as congestion pricing, ride-sharing, and traffic rationing, hold promise for mitigating capacity drop and improving the overall efficiency of traffic systems. 

\subsubsection{Traffic control}

\begin{figure}[!htbp]
\centering
\subfigure[Under demand variations \kern -0.6cm\label{fig:MFD_demands}]{
    {\includegraphics[width=0.45\linewidth]{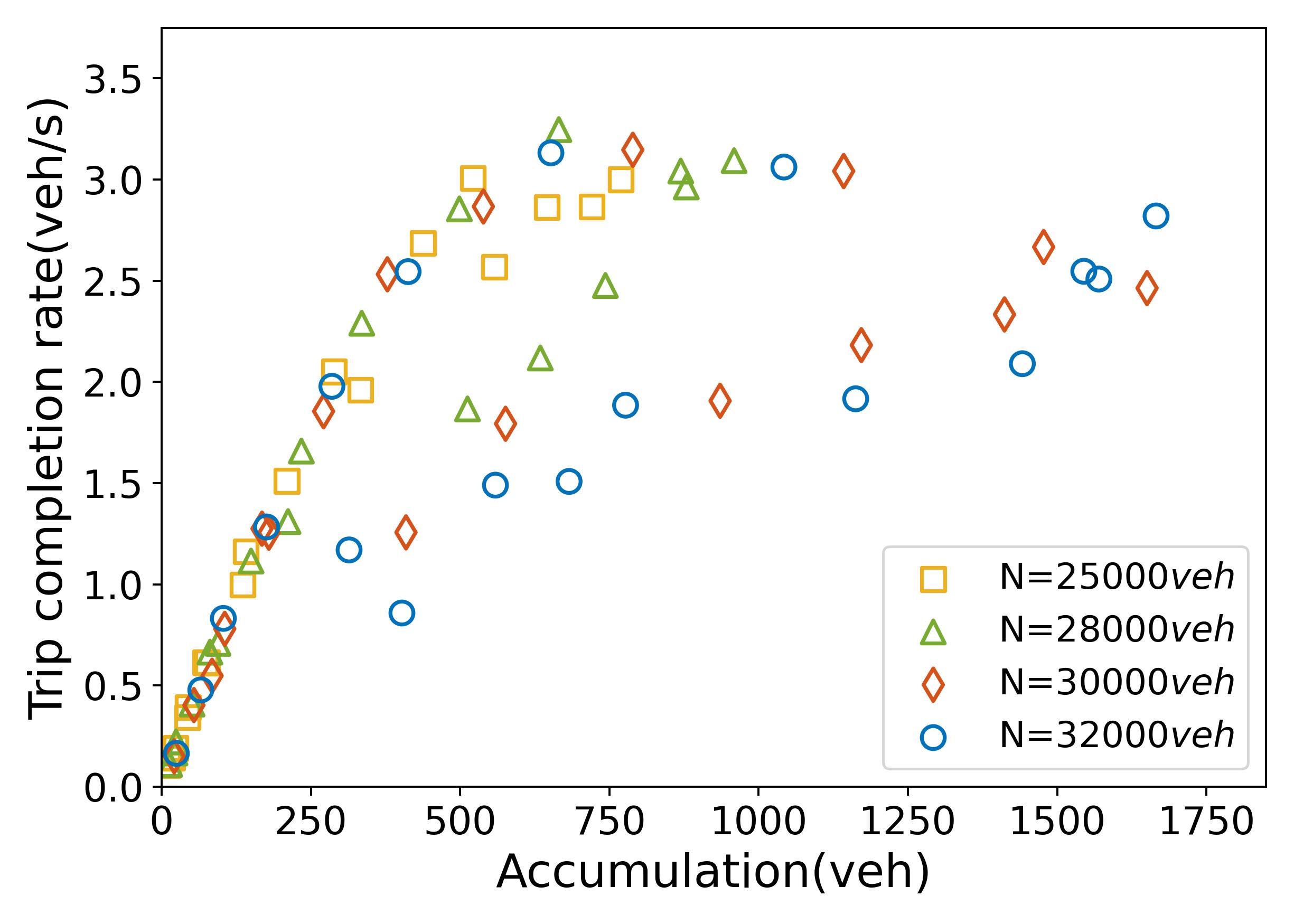}}
}
\subfigure[Under signal control variations \kern -0.6cm\label{fig:MFD_signal_control}]{
    {\includegraphics[width=0.45\linewidth]{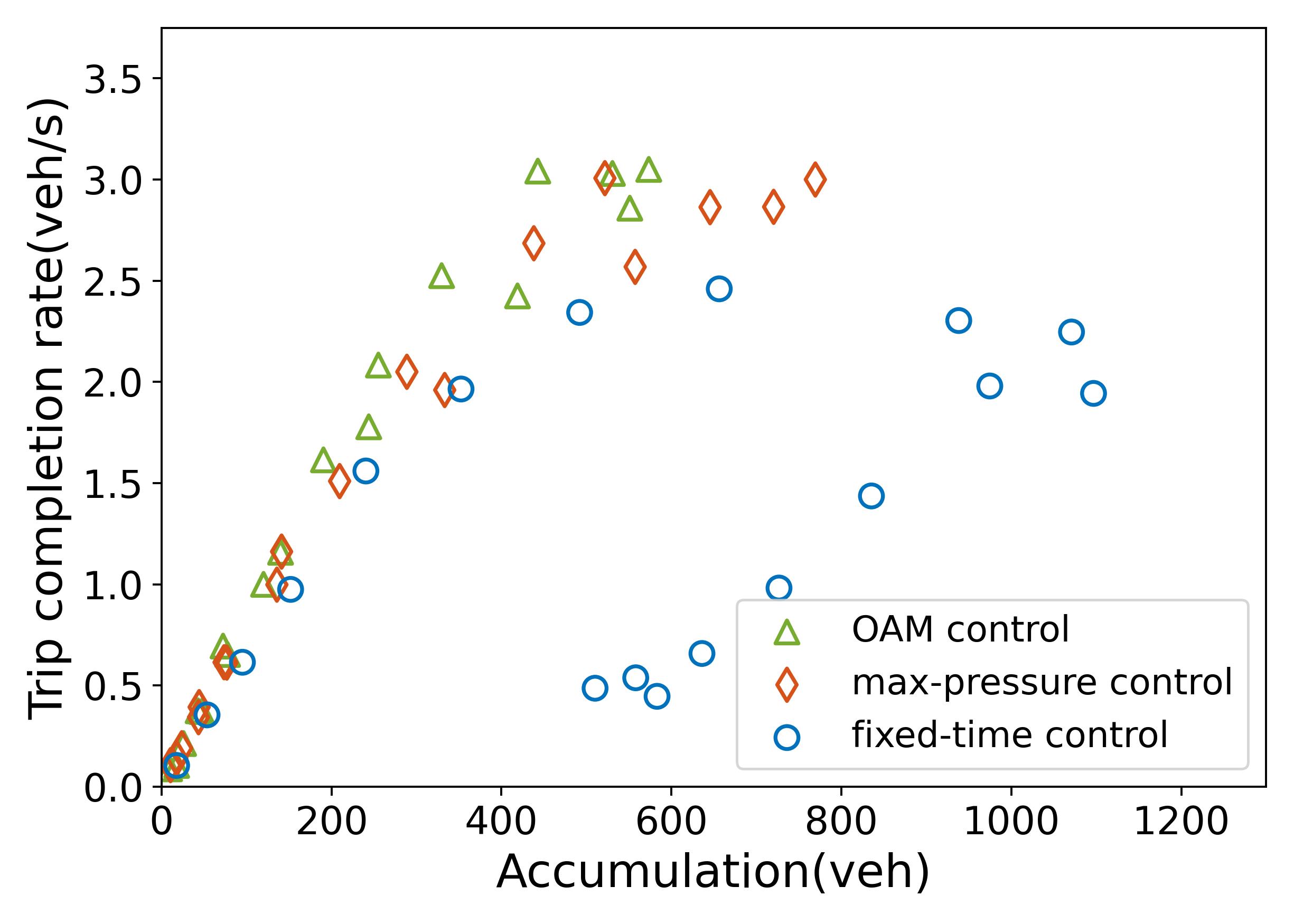}}
}
\subfigure[Under perimeter control variations \kern -0.6cm\label{fig:MFD_control}]{
    {\includegraphics[width=0.45\linewidth]{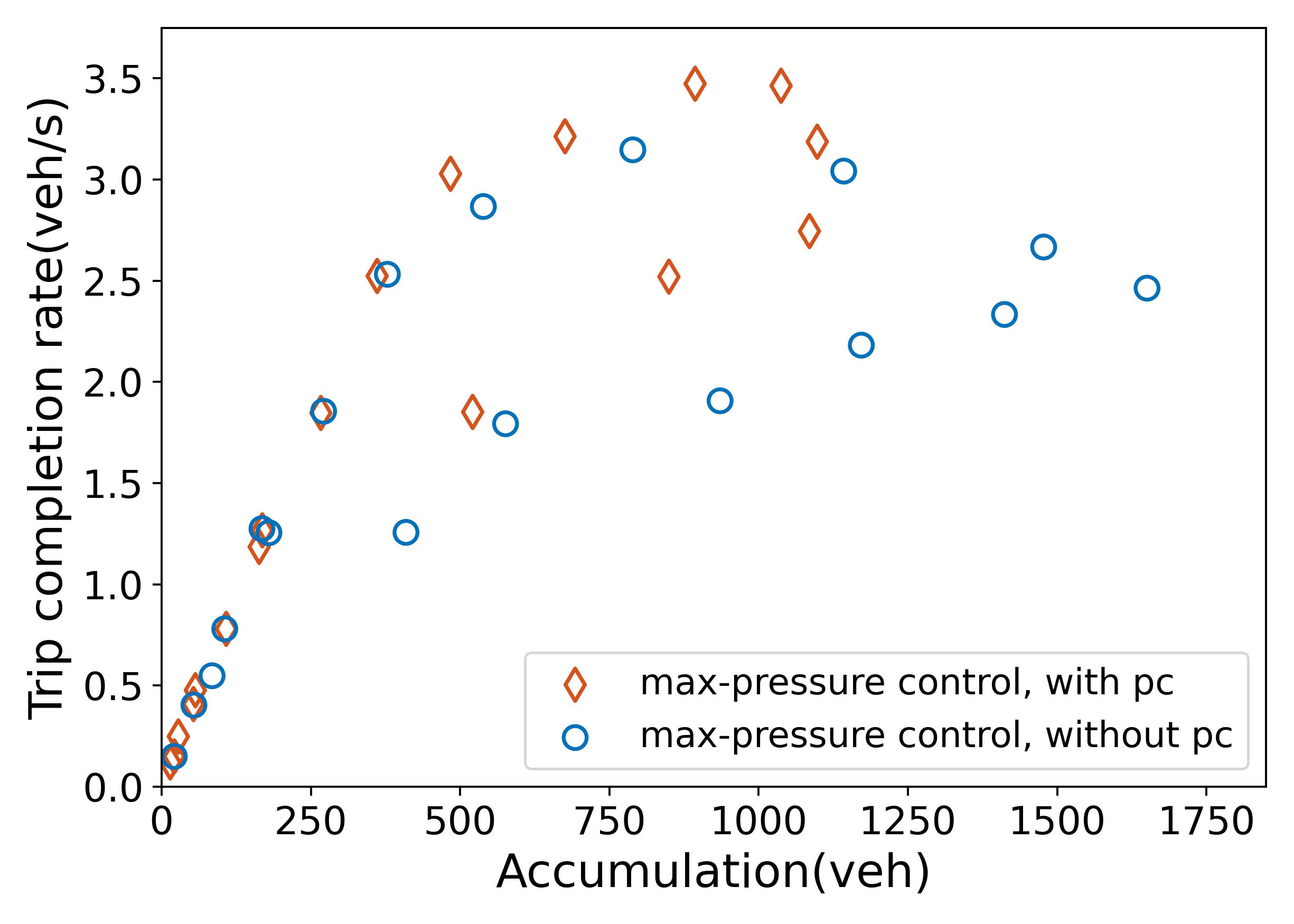}}
}
\subfigure[Under vehicle acceleration variations \kern -0.6cm\label{fig:MFD_acc}]{
    {\includegraphics[width=0.45\linewidth]{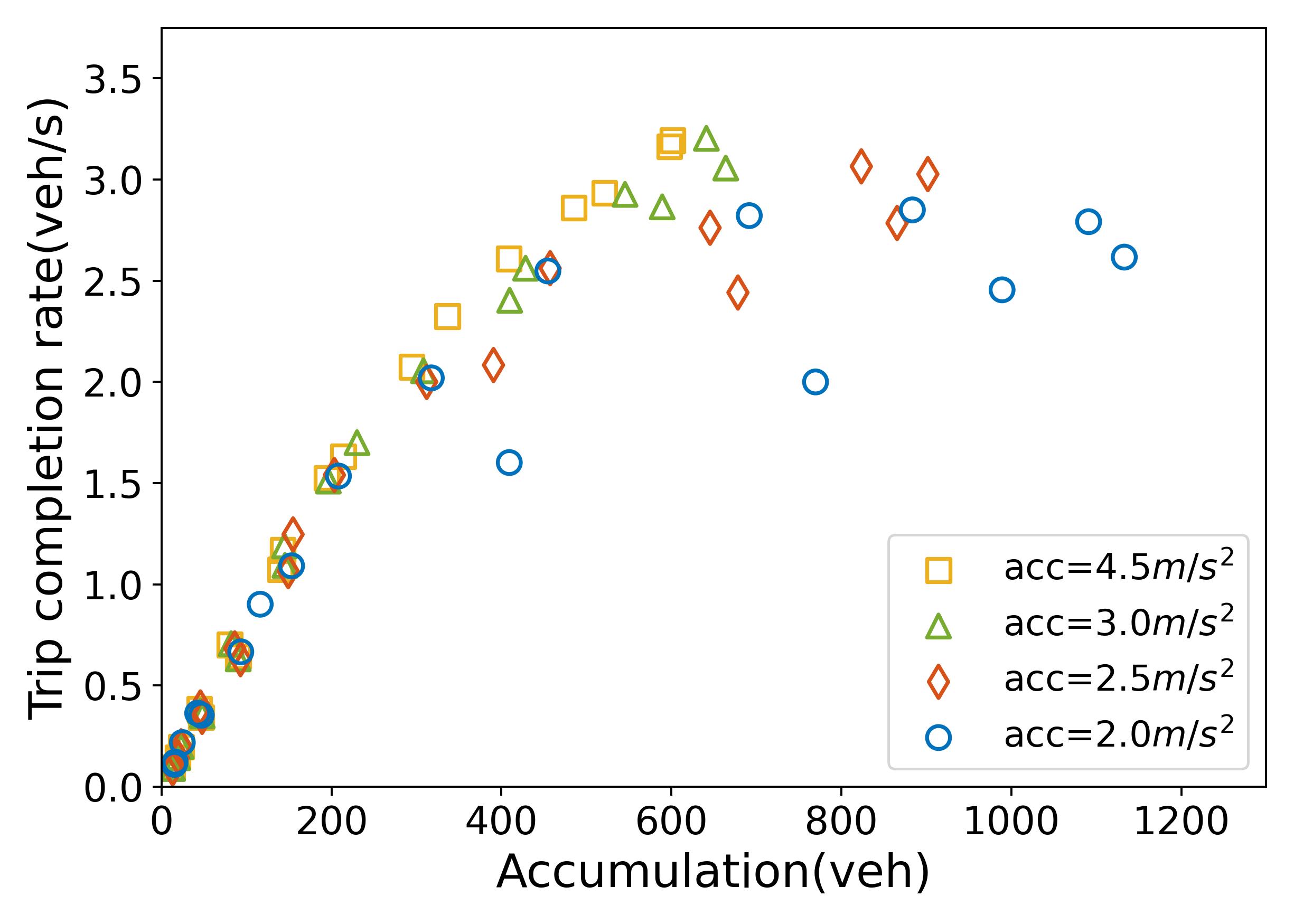}}
}
\caption{MFDs under different variations.}
\label{fig:simu_cali}
\end{figure}

In addition to travel demand, traffic control is another prominent factor influencing MFD hysteresis. 
We first analyze MFDs under various signal control strategies, including the fixed-time controller, max-pressure controller \citep{varaiya2013max, varaiya2013max2}, and the Option-Action Reinforcement Learning framework for universal Multi-intersection control (OAM) proposed in our previous work \citep{liang2022oam}.
Using a demand input $N$=25,000 $veh$ as an example, MFDs aggregated at 10-minute intervals under different signal control strategies are illustrated in \autoref{fig:MFD_signal_control}.

The fixed-time controller, which assigns equal durations to each phase, is overly simplistic and lacks the flexibility to adapt to fluctuating traffic volumes. Consequently, the MFD under this controller exhibits the largest hysteresis loop.
The maximal accumulation reaches 1100 $veh$, and the maximal trip completion rate is only about 2.4 $veh/s$. Moreover, network accumulation fails to fully dissolve within the simulation period.
The max-pressure controller, which adjusts phase durations based on the queues adjacent to the intersections, can better adapt to slow changes in demand patterns and maximize network throughput \citep{varaiya2013max2, varaiya2013max}. The MFD under max-pressure controller shows a reduced maximal accumulation of approximately 750 $veh$ compared to the fixed-time controller. Additionally, the maximal trip completion rate improves to about 2.95 $veh/s$, and the size of the hysteresis loop decreases significantly.
The OAM controller incorporates reinforcement learning to devise optimal signal plans while adapting to dynamic traffic congestion, providing strong generalization ability and outperforming existing signal control schemes \citep{liang2022oam}. 
The MFD under the OAM controller demonstrates further improvement, with reduced hysteresis size indicating its superiority in managing network traffic.
The maximal accumulation is about 580 $veh$, and the maximal trip completion rate is about 3.0 $veh/s$, the highest among these three cases.

Recent research reveals that traffic congestion can be greatly alleviated by identifying and regulating critical intersections within the network \citep{kouvelas2017enhancing, su2023hierarchical, tsitsokas2023two}. Perimeter control presents an effective and efficient way to improve network performance by controlling traffic flow at a macroscopic level, which can serve as a demand management scheme \citep{zhong2018boundary}.
We further examine the effect of perimeter control on MFD hysteresis based on the simulated network shown in \autoref{fig:topo_pc}.
We employ a perimeter control strategy on boundary roads following our previous work \citep{chen2022data}. For comparative analysis, we also implement a case with no perimeter control, i.e., control-free on boundary roads.

Taking the demand input $N$=30,000 $veh$ and intersections within the region deploying max-pressure controller as an example, MFDs aggregated at 10-minute intervals with and without perimeter control are shown in \autoref{fig:MFD_control}. 
A notable reduction in the size of the hysteresis loop is observed with the deployment of perimeter control. 
Without perimeter control on the boundary roads, vehicles with cross-boundary destinations access the other region directly, resulting in a sharp increase in the accumulation of the region. As shown in \autoref{fig:MFD_control}, the maximal accumulation reaches approximately 1600 $veh$, and the size of the hysteresis loop is almost doubled in the case with perimeter control.
The primary contribution of perimeter control is managing cross-boundary demand to prevent extreme congestion. Vehicles queue in a buffer zone until the perimeter control signal turns green, significantly reducing the maximal accumulation to only 1100 $veh$.
Additionally, the maximal trip completion rate without perimeter control is about 3.2 $veh/s$. Upon the implementation of perimeter control, the maximal trip completion rate improves notably to around 3.5 $veh/s$.

These results under different signal control and perimeter control schemes reveal that the deployment of well-designed control strategies can not only mitigate capacity drop but also significantly enhance traffic efficiency. 
A detailed microscopic simulation example based on perimeter control is provided in \autoref{Appendix:implications}, where the traffic efficiency represented by total time spent (TTS) and network accumulation evolutions are presented.
Furthermore, the findings highlight the importance of considering the ever-changing MFDs in the design of MFD-based traffic controllers. Besides, taking the interaction between MFD uncertainty and traffic control strategies into account may offer more valuable insights for traffic management.

\subsubsection{Vehicle acceleration and deceleration}

In addition to controllable factors such as travel demand and traffic control, MFD hysteresis is also influenced by uncontrollable factors, e.g., vehicle acceleration and deceleration. This factor is related to human driving behaviors, which are generally beyond the control of traffic systems.
To explore the effect of vehicle acceleration and deceleration on MFD hysteresis, we simulate scenarios with a demand input $N$=25,000 $veh$ to compare MFDs under varying vehicle accelerations ($m/s^2$), while maintaining a uniform deceleration of 4.5 $m/s^2$ in all cases. 

The resulting MFDs aggregated at 10-minute intervals are shown in \autoref{fig:MFD_acc}.
When vehicle acceleration is 4.5 $m/s^2$, equal to the vehicle deceleration, the hysteresis loop in MFD nearly disappears, demonstrating no significant reduction in trip completion rate during network recovery. 
The maximal accumulation is about 600 $veh$, and the maximal trip completion rate is about 3.2 $veh/s$. However, an acceleration speed equal to deceleration is ideal and rarely achievable in real-world urban traffic. 
When the acceleration is smaller than deceleration, e.g., 3.0 $m/s^2$, a slight hysteresis loop is observed, and the maximal accumulation increases to about 700 $veh$. 
With vehicle acceleration at 2.5 $m/s^2$, a much bigger hysteresis loop appears. The maximal accumulation grows to about 900 $veh$ and the maximal trip completion rate decreases to approximately 3.0 $veh/s$. 
Finally, when vehicle acceleration is 2.0 $m/s^2$, the hysteresis loop is roughly double the size of that in the 2.5 $m/s^2$ case. The maximal accumulation exceeds 1150 $veh$, and the maximal trip completion rate drops further to 2.8 $veh/s$.

These results show that the discrepancy between vehicle acceleration and deceleration contributes to more pronounced MFD hysteresis.
The decrease in vehicle acceleration leads to an extended duration for vehicles to attain target speeds, thereby increasing average travel time and reducing the maximal trip completion rate. The longer presence of vehicles worsens traffic congestion, which further increases the maximal accumulation of network.
Furthermore, the reduced vehicle acceleration significantly lengthens the recovery time of the network from congestion, resulting in a more pronounced hysteresis phenomenon.
However, managing individual driving behaviors remains a significant challenge. The future increasing penetration of connected automated vehicles can enable traffic managers to regulate the microscopic factors and reduce uncertainty caused by this discrepancy.

\subsection{Traffic resilience analysis related to MFD uncertainty}

Traffic resilience refers to the ability of traffic systems to adapt to changing conditions and withstand, respond to, and recover quickly from disruptions \citep{scope2015transportation}.
As stated in \cite{lu2024traffic}, traffic resilience loss occurs when traffic networks fail to operate optimally as accumulation exceeds the critical value. Additionally, the trip completion rate measured by MFDs serves as a measure of the service rate of the network.
Therefore, resilience loss can be evaluated by quantifying the reduction in trip completion rate at time $t$, which is calculated as follows \citep{lu2024traffic}:
\begin{align}
    \Delta(t)=\frac{G(t)-G_{max}}{G_{max}} \ T(n(t),n_{cr})
    \label{eq:resi_loss}
\end{align}
where $G_{max}$ is the maximal trip completion rate, and $n_{cr}$ is the critical network accumulation. $T(n(t),n_{cr})$ is an indicator function defined as follows:
\begin{align}
    T(n(t),n_{cr})=\left \{
    \begin{aligned}
        & 0, & n(t) < n_{cr};\\
        & 1, & n(t) \geq n_{cr}.
    \end{aligned}
    \right.
    \label{eq:resili_indicator}
\end{align}

\begin{figure}[!htbp]
\centering
\subfigure[Under demand variations \kern -1cm]{ \label{fig:resi_demand}
    \includegraphics[width=0.45\linewidth]{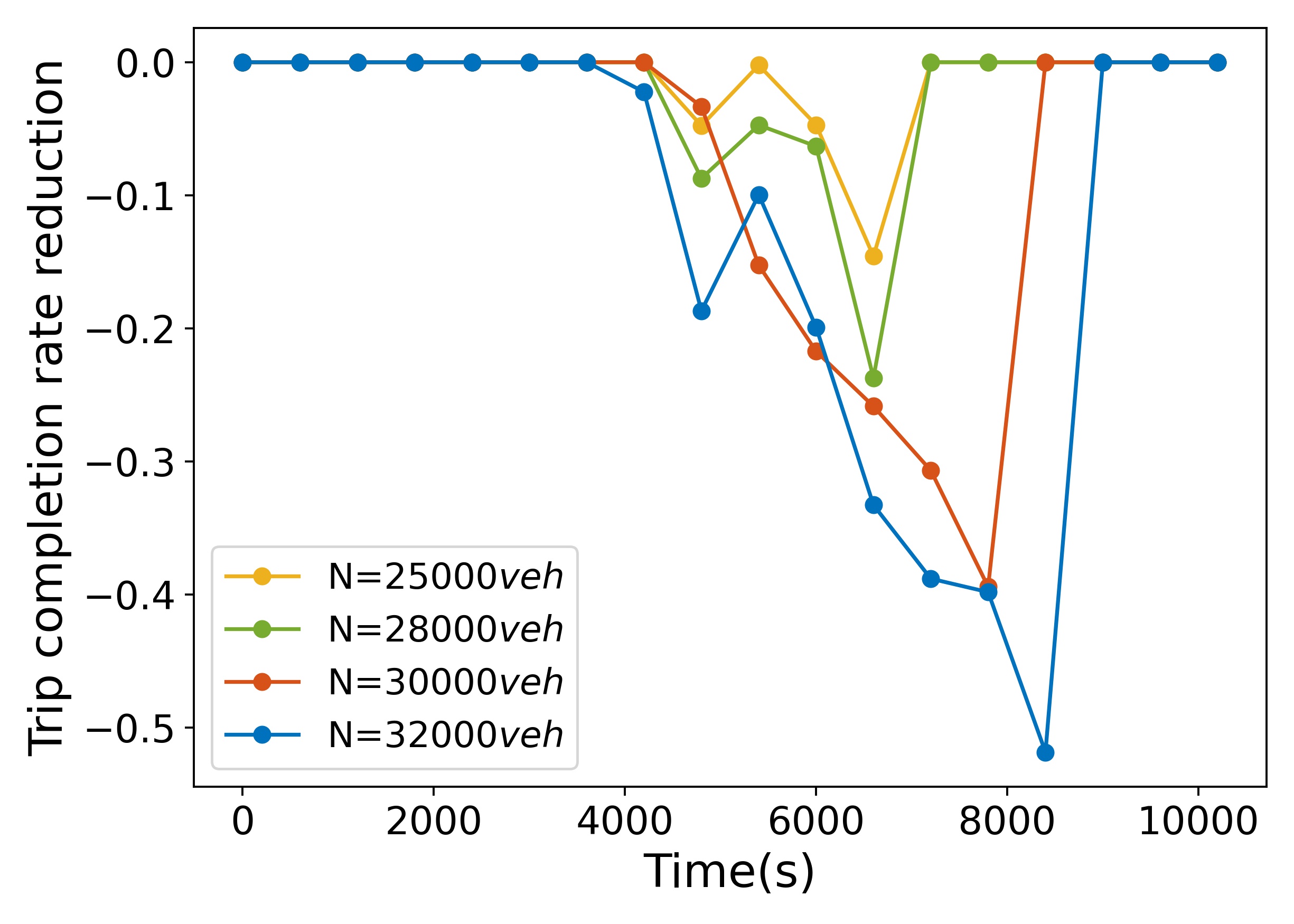}
    }
\subfigure[Under signal control variations \kern -1cm]{ \label{fig:resi_signal_ctrl}
    \includegraphics[width=0.45\linewidth]{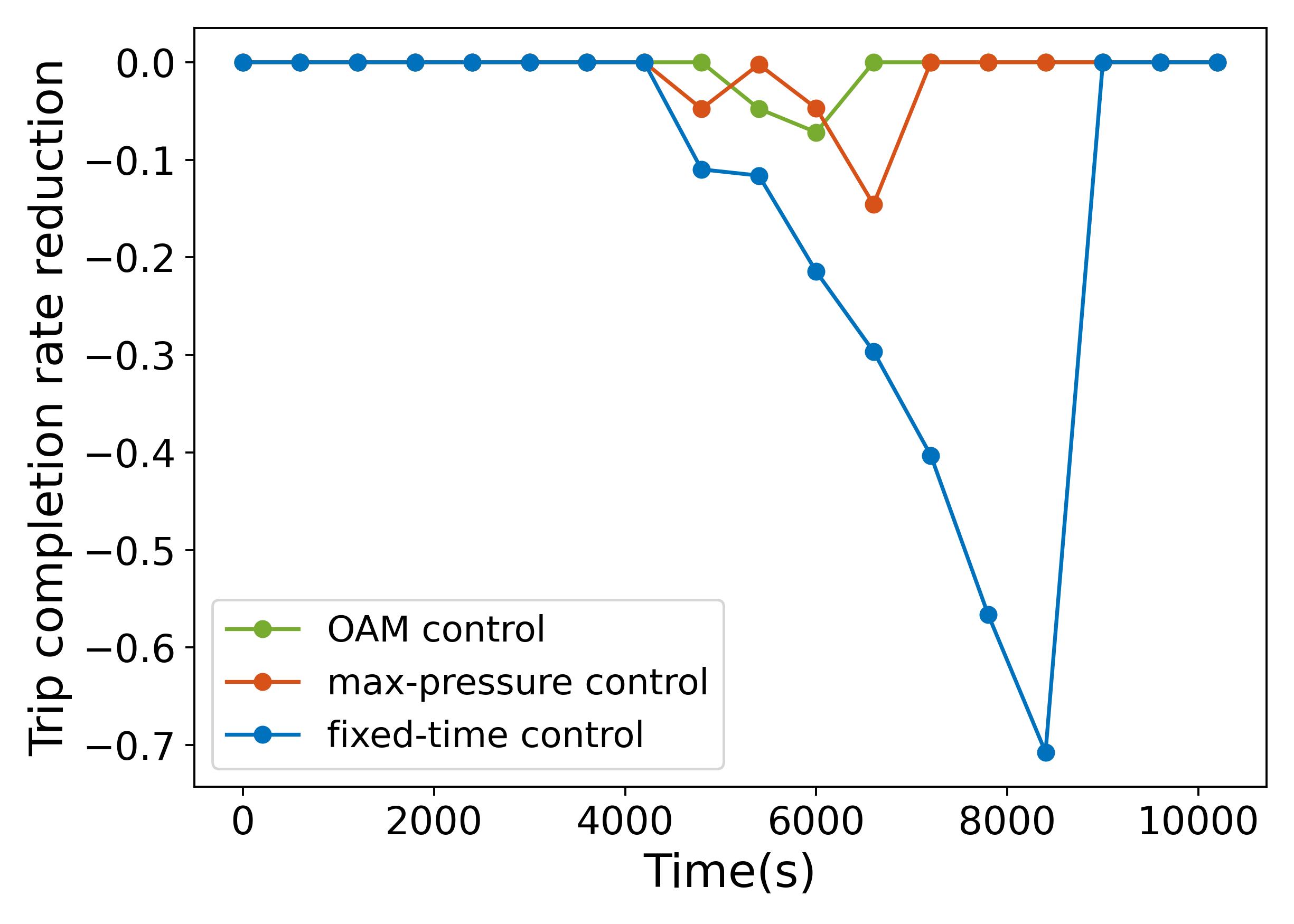}
    }
\subfigure[Under perimeter control variations \kern -1cm]{ \label{fig:resi_ctrl}
    \includegraphics[width=0.45\linewidth]{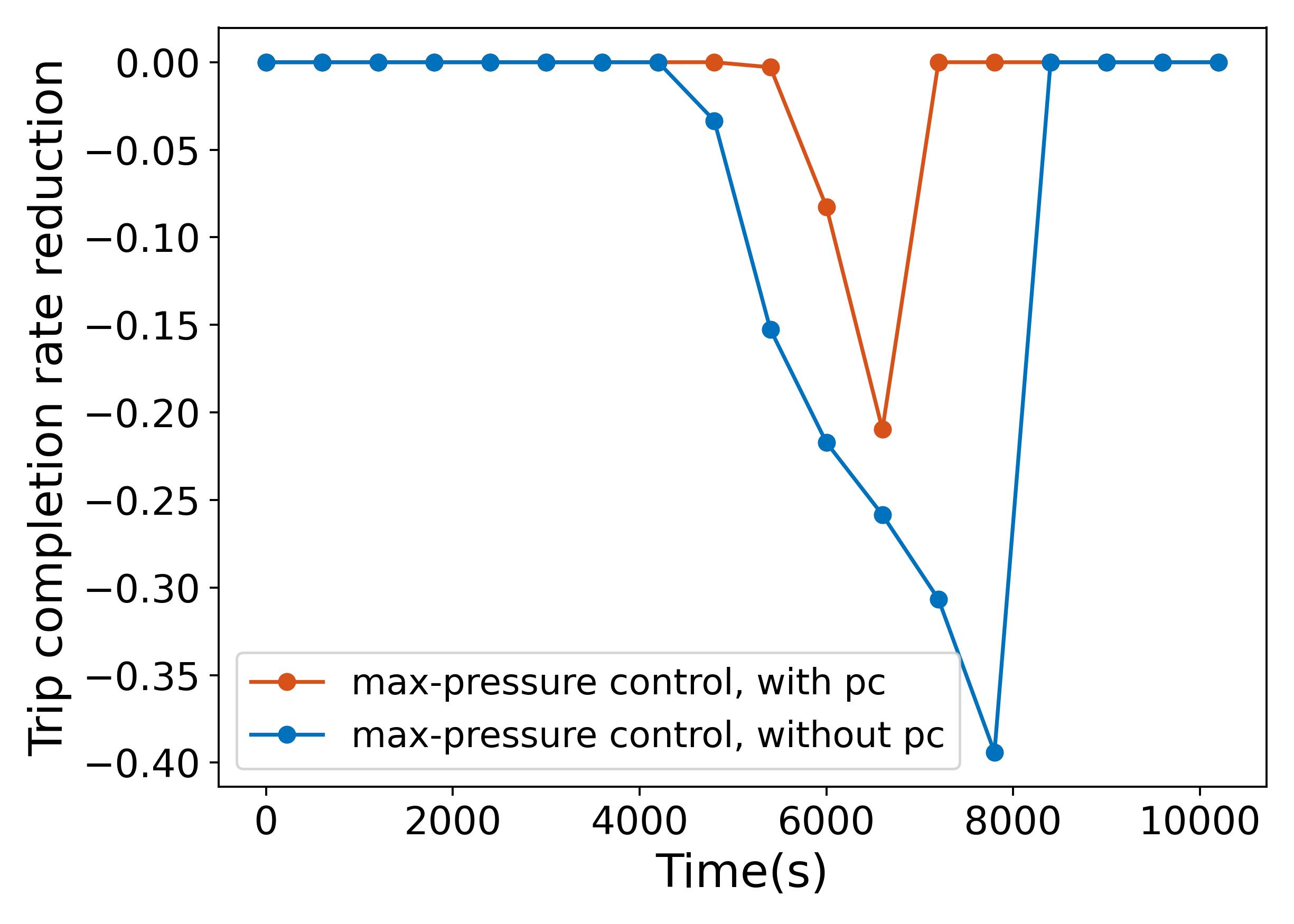}
    }
\subfigure[Under vehicle acceleration variations \kern -1cm]{ \label{fig:resi_acc}
    \includegraphics[width=0.45\linewidth]{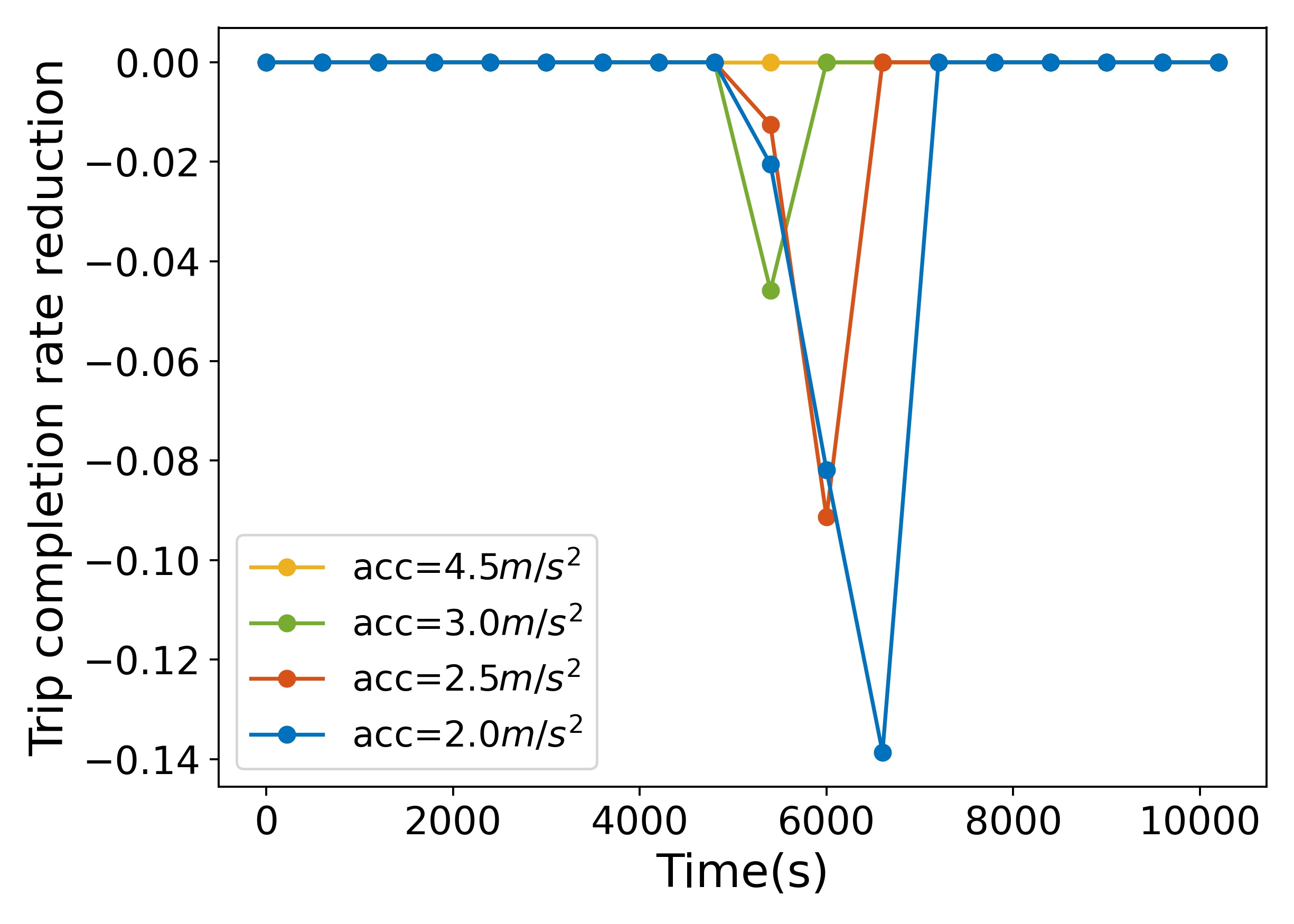}
    }
\caption{Traffic resilience under different variations.}
\label{fig:resilience_cal}
\end{figure}

The resilience loss curves under different demand inputs are shown in \autoref{fig:resi_demand}, where the loss is calculated at 10-minute intervals.
It is evident that traffic systems exhibit greater resilience under lower demand inputs, represented by a smaller area enclosed by the ``resilience triangle".
As previously mentioned, the hysteresis loop occurs when the traffic network recovers from extreme congestion \citep{li2023perimeter}.
Under lower demand inputs, as depicted in \autoref{fig:MFD_demands}, the hysteresis loop is relatively small, and the trip completion rate experiences only a slight loss during the recovery phase. Besides, the loss does not last a long time and recovers to its optimal service rate immediately. On the contrary, for the higher demand input cases, the trip completion rate loss persists for a longer duration, indicating a prolonged recovery time to attain the optimal service rate, signifying poorer traffic resilience.

Likewise, a traffic system equipped with an efficient signal control scheme or perimeter control strategy exhibits greater resilience, as illustrated in  \autoref{fig:resi_signal_ctrl} and \autoref{fig:resi_ctrl}.
As shown in \autoref{fig:MFD_signal_control}, when the traffic system reaches the congestion state, a noticeable loss occurs and persists for an extended period before recovering to its optimal service rate under the fixed-time controller. Nevertheless, when an efficient control strategy (e.g., max-pressure or OAM) is deployed, network traffic is better managed and prevented from getting into severe congestion. Although the trip completion rate loss still occurs, the recovery time is notably shorter. 
The same reasoning applies to the scenarios under perimeter control variations as shown in \autoref{fig:MFD_control}. Both findings underscore the effectiveness of well-designed control strategies in enhancing traffic resilience.

Furthermore, resilience loss under vehicle acceleration variations is shown in \autoref{fig:resi_acc}, indicating that the increased discrepancy between vehicle acceleration and deceleration worsens traffic resilience. 
As evidenced in \autoref{fig:MFD_acc}, the hysteresis loop nearly disappears when vehicle acceleration equals deceleration. The corresponding resilience loss is zero, signifying that the system operates optimally throughout the simulation period. Conversely, as vehicle acceleration decreases, the recovery time of resilience loss lengthens, and the minimum value deteriorates.
These results strongly support that heterogeneous driving behaviors, e.g., driving with lower acceleration, reduce traffic resilience and bring more uncertainty to the traffic system.

In summary, MFD hysteresis provides an intuitive tool to characterize traffic system resilience in the context of travel demand, traffic control, and vehicle acceleration variations. Additionally, it offers a straightforward way to represent the relationship between MFD uncertainty and network performance, thereby improving the understanding of traffic system responses and recovery from various disruptions.

\section{Conclusions} \label{sec:MFD:cali:conclusion} 
Traffic congestion remains a persistent issue arising from the imbalance between travel demand and network capacity, necessitating a thorough understanding of network capacity for effective traffic control and management. 
MFD provides an efficient framework for quantifying network capacity, offering insights into the network supply function. However, MFDs derived from empirical or simulated data are rarely well-defined and instead exhibit significant scatter and uncertainty. 
To facilitate traffic management and control design, it is essential to quantify the uncertainty within MFDs to capture traffic dynamics accurately.
This paper first proposed a unified mathematical program for simultaneous MFD calibration and uncertainty quantification.
Subsequently, conventional quantification principles were included as special cases to develop two approaches for capturing MFD data scatter, and further evaluated using data from two urban networks in China.
Both approaches quantified the scatter width by measuring data distributions between different calibrated curves. 
However, these approaches did not provide clear guidance on selecting representative curves to quantify the MFD uncertainty. In addition, the resulting curves lacked physical interpretability, limiting their applicability in traffic control and demand management design.

Motivated by the observed MFD hysteresis phenomenon, this paper inferred that different physics of congestion loading and recovery phases contribute to the data scatter and developed a novel uncertainty quantification approach considering distinct congestion phases with parameter $\delta$.
Microscopic simulation results validated the effectiveness of the proposed quantification approach in capturing congestion loading and recovery phases. 
The parameter $\delta$ can be interpreted as an indicator of the distance between congestion loading and recovery, corresponding to the capacity drop.
Moreover, $\delta$ provides a potential way for incorporating MFD uncertainty into traffic control and management design processes. For example, it can be integrated into objective functions or used as a regularization term in traffic control optimization.
Traffic control strategies tailored to distinct congestion phases based on $\delta$ would be better suited for managing real-world traffic conditions. 


Furthermore, we investigated the effects of macroscopic and microscopic factors contributing to network capacity and MFD uncertainty. 
Simulation results revealed that the MFD hysteresis loop increased with higher demand input, and well-designed control strategies better managed network traffic and prevented severe congestion.
These findings suggest that controllable factors, including travel demand and traffic control, enable the development of both demand management and control strategies to improve traffic efficiency. 
On a microscopic level, increased discrepancies between vehicle acceleration and deceleration amplified MFD hysteresis. However, managing individual driving behaviors remains challenging until the widespread adoption of connected automated vehicles.
Finally, we employed an MFD-based traffic resilience indicator to explore the connection between MFD uncertainty and network performance. The findings highlighted that traffic systems with lower MFD uncertainty exhibit greater resilience in responding to congestion disruptions.

While this paper has made a contribution to the uncertainty quantification of MFDs, several directions remain for future research.
The first is the refinement of the quantification approach. Future work could focus on enhancing the approach to account for additional factors, such as environmental conditions, weather, and so on.
Additionally, exploring adaptive traffic control strategies tailored to varying sources and levels of MFD uncertainty could be another interesting direction.
The development of such control strategies holds the potential to enhance the efficiency and resilience of urban traffic systems.

\section*{Acknowledgments}

Financial support from the National Natural Science Foundation of China (No. 72071214) is gratefully acknowledged.


\appendix

\section{Scenario settings for influencing factors analysis} \label{appendix_setting}
\begin{table}[!ht]
\centering
\caption{Scenario settings for influencing factors analysis}
\label{tab:scenario_setting}
\begin{tabular}{p{3.0cm} p{2.5cm} p{3.5cm} c c}
\hline
 & \centering Demand inputs ($veh$) & \centering Vehicle acceleration ($m/s^2$) & Signal control & Perimeter control \\ \hline
\multirow{4}{3.2cm}{Demand variations} 
& \centering 25,000 & \centering 3.0 & max-pressure & $\times$ \\
& \centering 28,000 & \centering 3.0 & max-pressure  & $\times$ \\
& \centering 30,000 & \centering 3.0 & max-pressure  & $\times$ \\
& \centering 32,000 & \centering 3.0 & max-pressure  & $\times$ \\ \hline
\multirow{3}{3.2cm}{Signal control variations} 
& \centering 25,000 & \centering 3.0 & fixed-time & $\times$ \\
& \centering 25,000 & \centering 3.0 & max-pressure & $\times$ \\
& \centering 25,000 & \centering 3.0 & OAM & $\times$ \\ \hline
\multirow{2}{3.2cm}{Perimeter control variations} 
& \centering 30,000 & \centering 3.0 & max-pressure  & $\times$ \\
& \centering 30,000 & \centering 3.0 & max-pressure  & $\checkmark$ \\ \hline
\multirow{4}{3.2cm}{Vehicle acceleration variations} 
& \centering 25,000 & \centering 4.5 & max-pressure  & $\checkmark$ \\ 
& \centering 25,000 & \centering 3.0 & max-pressure  & $\checkmark$ \\ 
& \centering 25,000 & \centering 2.5 & max-pressure  & $\checkmark$ \\ 
& \centering 25,000 & \centering 2.0 & max-pressure  & $\checkmark$ \\
\hline
\end{tabular}
\\\raggedright
\footnotesize{\textit{Note:} 
This table presents the scenario settings. Specifically, signal control is deployed within the region at internal intersections, and perimeter control is applied at boundary intersections. }
\end{table}

\section{A microscopic simulation example} \label{Appendix:implications}

\begin{figure}[!htbp]
\centering
\subfigure[The time-varying demand input \kern -0.8cm \label{fig:demand_input}]{
    {\includegraphics[width=0.45\linewidth]{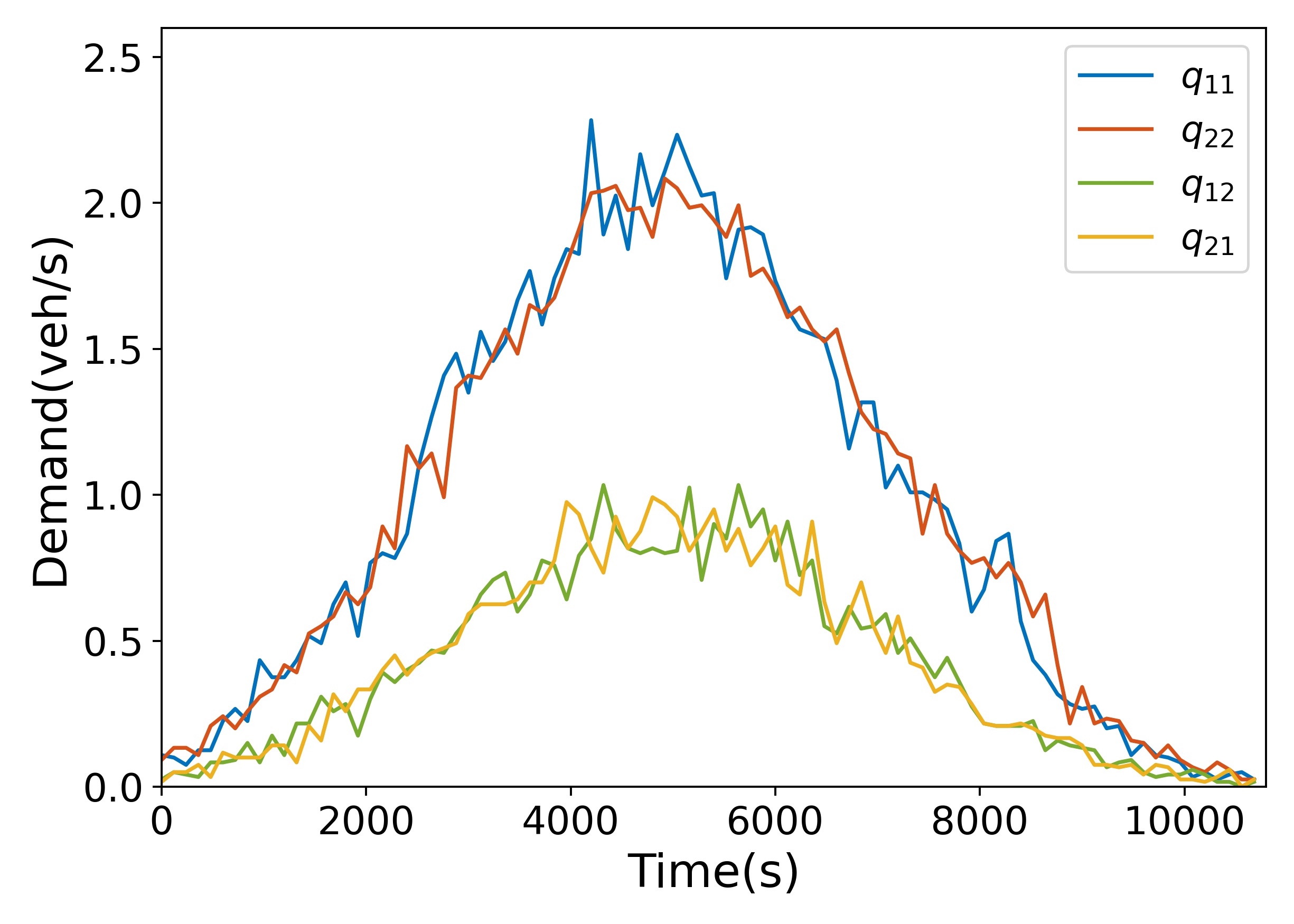}}
}
\subfigure[Comparison of total time spent (TTS) \kern -0.8cm\label{fig:TTS_vs}]{
    {\includegraphics[width=0.45\linewidth]{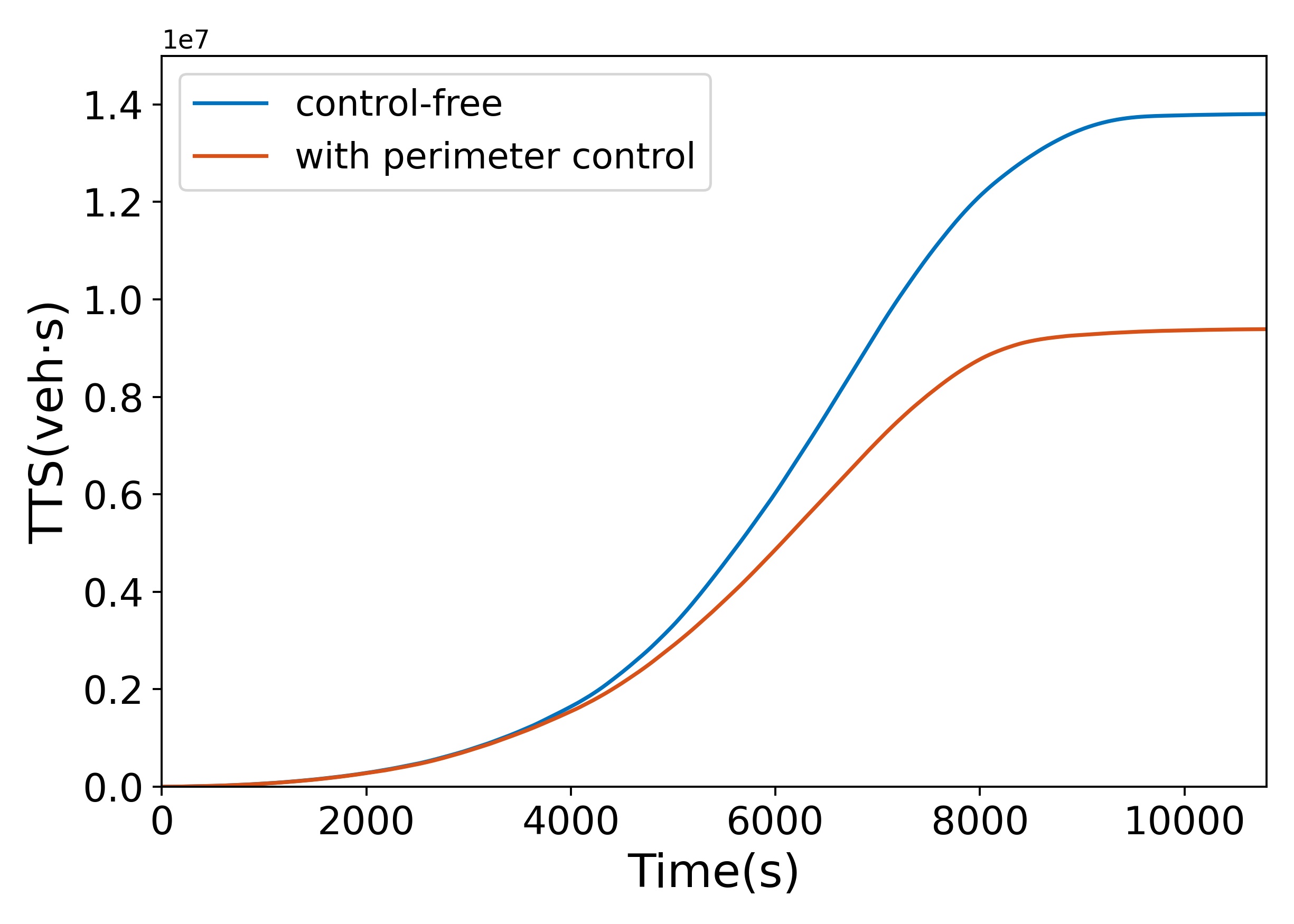}}
}
\subfigure[Comparison of MFDs for Region 1 \kern -0.8cm\label{fig:MFD_iter_R1}]{
    {\includegraphics[width=0.45\linewidth]{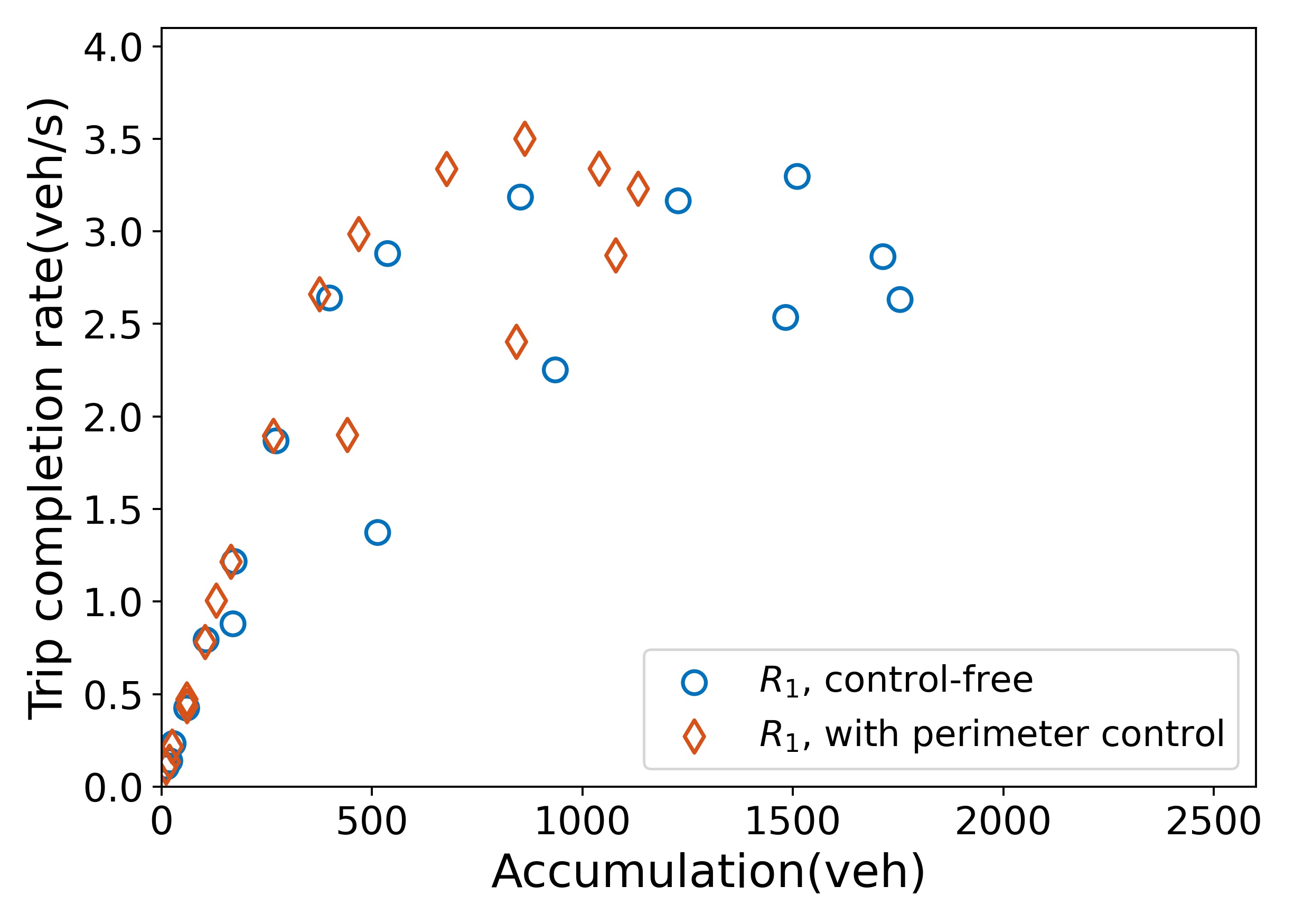}}
}
\subfigure[Comparison of MFDs for Region 2 \kern -0.8cm\label{fig:MFD_iter_R2}]{
    {\includegraphics[width=0.45\linewidth]{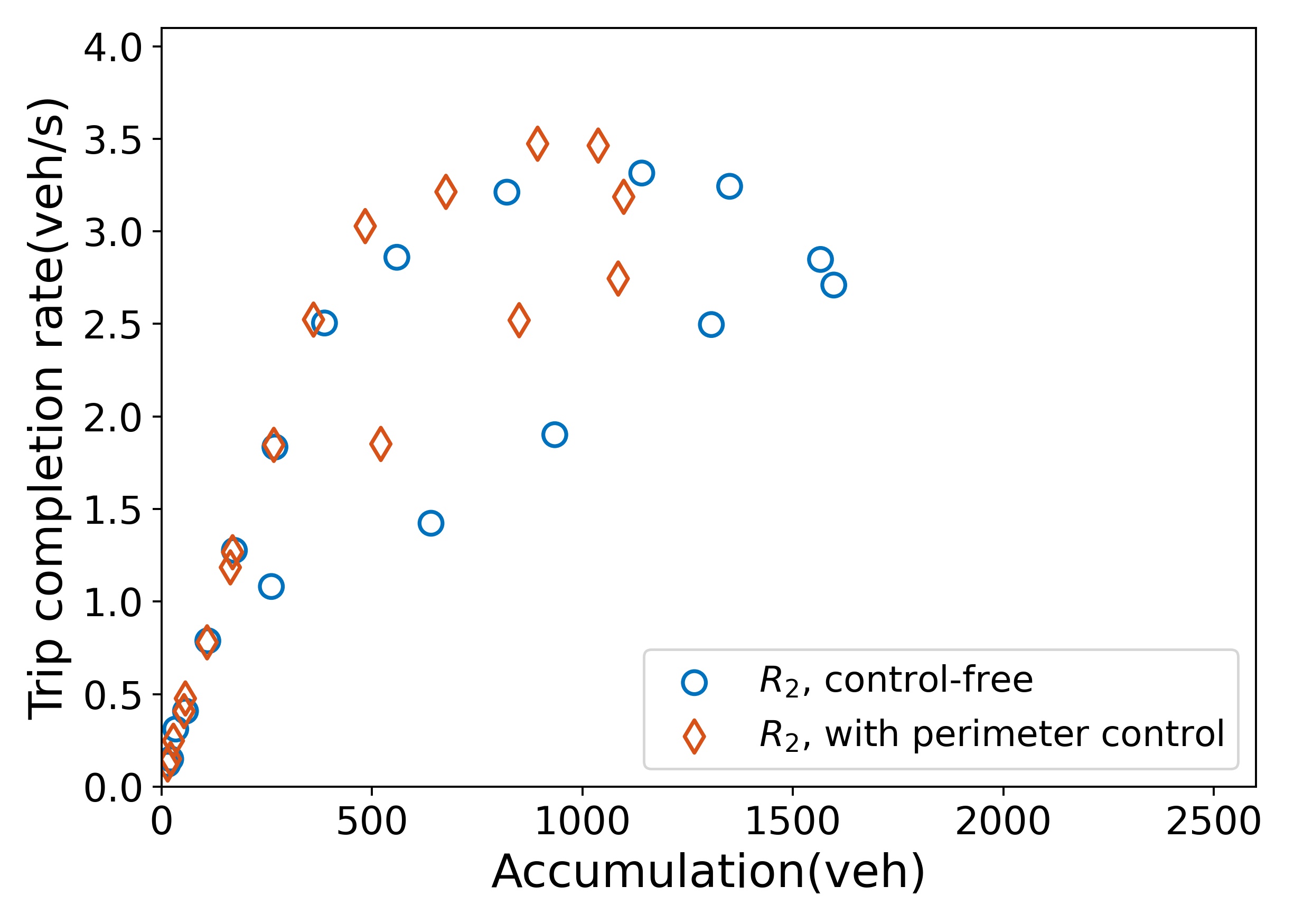}}
}
\subfigure[Network accumulation evolution for Region 1 \kern -0.8cm\label{fig:n_iter_R1}]{
    {\includegraphics[width=0.45\linewidth]{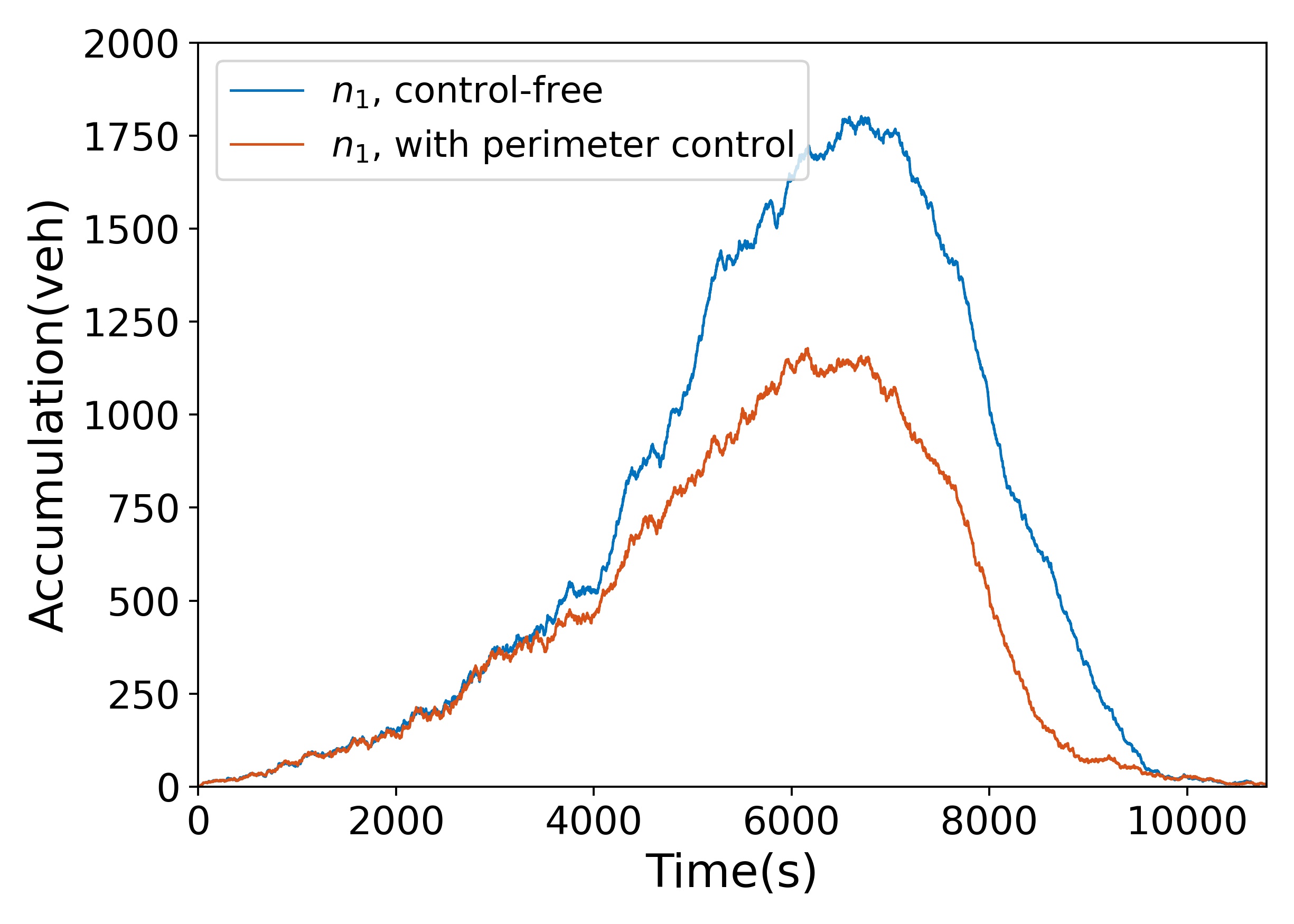}}
}
\subfigure[Network accumulation evolution for Region 2 \kern -0.8cm\label{fig:n_iter_R2}]{
    {\includegraphics[width=0.45\linewidth]{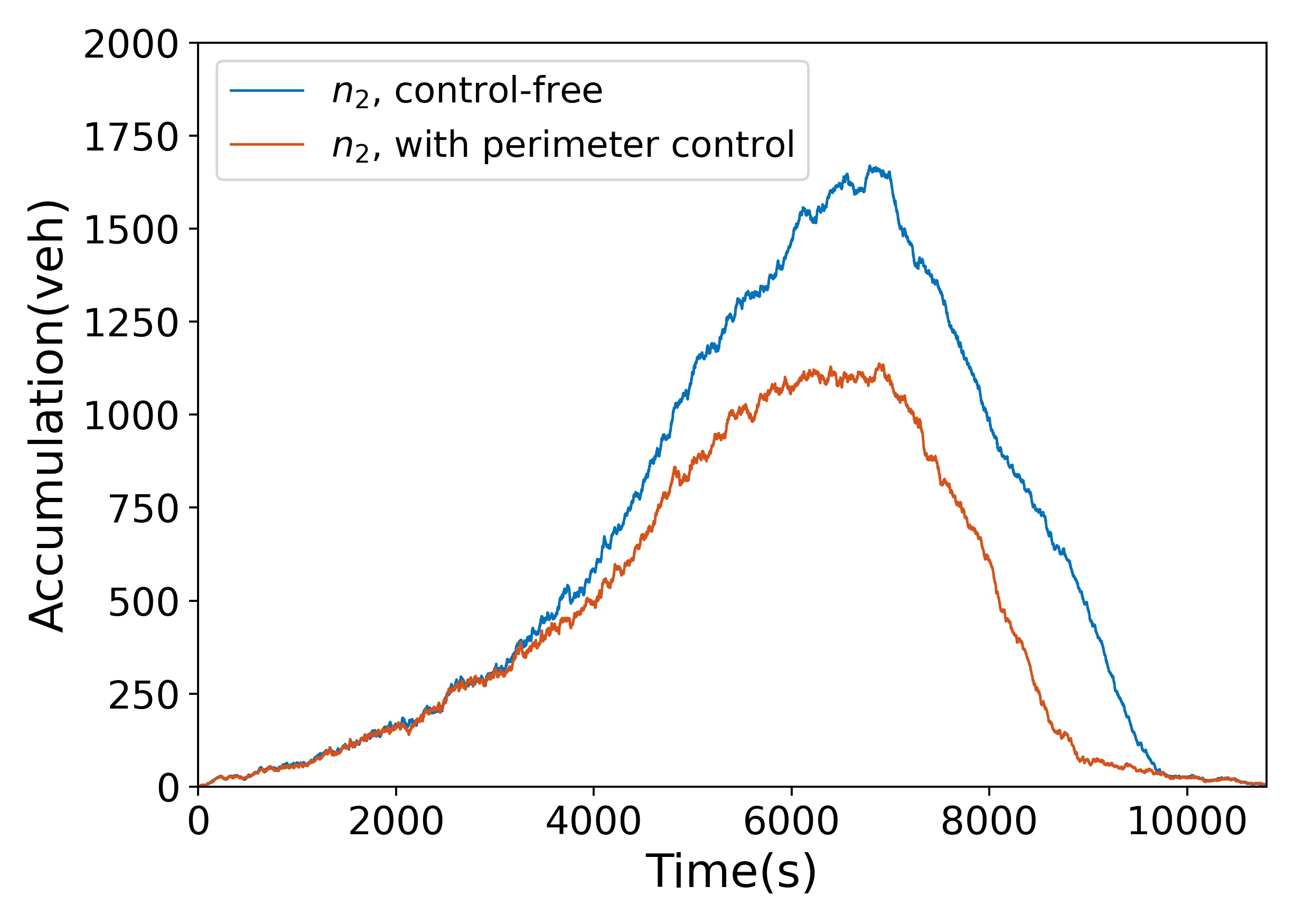}}
}
\caption{Microscopic simulation results.} 
\label{fig:simu_PC_implication}
\end{figure}

To further explore the implications of uncertainty quantification for dynamic traffic control and management, we conduct a simulated experiment using the microscopic network shown in \autoref{fig:topo_pc}.  
Given the effectiveness of perimeter control in regulating travel demand into a region, we examine the network performance under scenarios with and without perimeter control.
\autoref{fig:demand_input} shows the time-varying demand input for the network, with a total demand of $N=30,000$ $veh$. 
The baseline scenario is control-free, i.e., without perimeter control on boundary roads.
The total time spent (TTS) comparison is shown in \autoref{fig:TTS_vs}, which decreases by about 35\% after the deployment of the perimeter controller.
Besides, the MFDs of two regions, presented in \autoref{fig:MFD_iter_R1} and \autoref{fig:MFD_iter_R2}, show a significant reduction in MFD hysteresis, highlighting the potential of demand management strategies in mitigating MFD uncertainty. Moreover, the deployment of the controller results in a higher maximal trip completion rate, enabling the network to transition into the recovery phase earlier rather than fall into extreme congestion.
Furthermore, the accumulation evolutions are depicted in \autoref{fig:n_iter_R1} and \autoref{fig:n_iter_R2}.
The results indicate that perimeter control can effectively regulate accumulation below a lower level, thereby significantly mitigating congestion.
These findings validate the effectiveness of demand management strategies, such as perimeter control, in regulating traffic flow and reducing MFD uncertainty.

\normalem 
\bibliography{main}
\bibliographystyle{apalike}

\end{sloppypar}
\end{document}